\documentclass[12pt,preprint]{aastex}

\def\gtapr {\lower .1ex\hbox{\rlap{\raise .6ex\hbox{\hskip .3ex
        {\ifmmode{\scriptscriptstyle >}\else
                {$\scriptscriptstyle >$}\fi}}}
        \kern -.4ex{\ifmmode{\scriptscriptstyle \sim}\else
                {$\scriptscriptstyle\sim$}\fi}}}
\def \fred{\hbox{$f_{red}$}}
\def \sig5{\hbox{$\Sigma_5$}}
\def \rcl{\hbox{$r_{CL}$}}
\def \msun{${\rm M}_\odot$}
\shortauthors{Li et al.}
\shorttitle{CNOC1 Clusters}

\begin{document}

\title{Individual and Group Galaxies in CNOC1 Clusters}

\author{I.H. Li\altaffilmark{1,2}}
\email{tli@astro.swin.edu.au}
\author{H.K.C. Yee\altaffilmark{1}}
\email{hyee@astro.utoronto.ca}
\author{E. Ellingson\altaffilmark{3}}
\email{elling@casa.colorado.edu}
\altaffiltext{1}{Department of Astronomy and Astrophysics, University of Toronto, 50 St.~George Street, Toronto, ON, Canada, M5S 3H4}
\altaffiltext{2}{current address: Centre for Astrophysics and Supercomputing, Swinburne University, Hawthorn, Victoria, Australia, 3122}
\altaffiltext{3}{Center for Astrophysics and Space Astronomy, 389UCB, University of Colorado, Boulder, CO 80309}

\begin{abstract}
	Using wide-field $BVR_cI$ imaging for a sample of 16 intermediate redshift ($0.17 < z < 0.55$) galaxy clusters from the Canadian Network for 
Observational Cosmology (CNOC1) Survey, we investigate the dependence of 
cluster galaxy populations and their evolution on environment.
 Galaxy photometric redshifts are estimated using an empirical photometric redshift technique and galaxy groups are identified using a modified friends-of-friends algorithm in photometric redshift space. 
We utilize the red galaxy fraction ($f_{red}$) to infer the evolutionary status of 
galaxies in clusters, using both individual galaxies and galaxies in groups.
We apply the local galaxy density, \sig5, derived using the 
fifth nearest-neighbor distance, as a measure of local environment, and the cluster-centric radius, \rcl, as a proxy for global cluster environment.
Our cluster sample exhibits a Butcher-Oemler effect in both luminosity-selected
and stellar-mass-selected samples. 
We find that \fred~depends strongly on \sig5~and \rcl, and the Butcher-Oemler effect is observed in all \sig5~and \rcl~bins.
However, when the cluster galaxies are separated into \rcl~bins, or
into group and non-group subsamples, the dependence on local galaxy 
density becomes much weaker.
This suggests that the properties of the dark matter halo in which the
galaxy resides have a dominant effect on its galaxy population
 and evolutionary history.
We find that our data are consistent with the scenario that
cluster galaxies situated in successively richer groups
(i.e., more massive dark matter halos) reach a high \fred~value at 
earlier redshifts.
Associated with this,
we observe a clear signature of `pre-processing', in which cluster galaxies
belonging to moderately massive infalling galaxy groups 
show a much stronger evolution in \fred~than those classified
as non-group galaxies, especially at the outskirts of the cluster.
This result suggests that galaxies in groups infalling into clusters
 are significant contributors to the Butcher-Oemler effect.

\end{abstract}

\section{Introduction}
	With increasingly detailed studies of galaxy properties such as stellar population and morphology, our knowledge of galaxies in the Universe has greatly improved. 
The large-scale structure in the Universe shows that some galaxies are located in low-density filaments while others populate high-density clusters. 
This complicates our understanding of galaxy evolution over the Hubble time, because galaxy properties depend not only on time but also on environment. 
Therefore, in order to understand galaxy evolution in the Universe, galaxy properties require viewing simultaneously and systematically over time and in different environments.

	The morphology-density relation \citep{1980ApJS...42..565D} is well-accepted evidence for the environmental dependence of galaxy properties.
 Early-type galaxies favor dense environments, while late-type galaxies are common in less dense environments. 
Following this pioneering work, a considerable number of studies on this topic have been carried out or extended \citep[e.g.,][]{2002MNRAS.335..825D, 2003MNRAS.346..601G, 2003ApJ...585L...5H, 2005MNRAS.362..268T, 2005ApJ...626L..77N}, showing that environment is an important predictor of galaxy properties.

	Abundant in galaxy aggregations, clusters of galaxies offer the best sites to study such environmental effects. 
The observation that clusters at higher redshift contain a larger portion of blue galaxies is commonly referred to as the \it Butcher-Oemler effect \rm \citep[e.g.,][]{1984ApJ...285..426B,2006ApJ...642..188P,2001ApJ...547..609E,2007MNRAS.tmp..222G}.
The increase by $\sim20\%$ in the blue galaxy fraction from $z$=0.03 to $z$=0.50 in the Butcher-Oemler effect suggests that star forming activity in galaxy clusters decreases toward the nearby Universe. 
With the continuous accretion of field galaxies into galaxy clusters, this decrease implies that star formation is inhibited by the cluster environment. 

	However, the fraction of galaxies located in these densest galaxy environments is only $\sim10\%$ even at the present epoch. 
Groups of galaxies, the intermediate density locations between dense clusters and the sparse field, 
are the most common environment where galaxies are found. 
Seeded by lower-density perturbations in a hierarchical Universe, galaxy groups have a large variety of galaxy densities, 
from early-collapsed `fossil~groups' to aggregations of a few galaxies with densities only slightly greater than that of the field. 
With relatively small velocity dispersions compared with clusters, galaxy groups are also the favored environment for slow dynamical encounters, leading to mergers \citep[e.g.,][]{2000ASPC..197..377M,2007AJ....133.2630C}. 
Furthermore, a higher fraction of passive galaxies have been observed in galaxy groups than in the field \citep[e.g.,][]{2005MNRAS.358...71W,2008ApJ...680.1009W,2008arXiv0805.0004Y}.
A sort of morphology segregation in galaxy groups is observed in the nearby Universe, as early-type galaxies are more commonly located in the group centers and spirals have larger projected centric distance from the host groups \citep[e.g.,][]{2007ApJ...658..865J,2008A&A...486....9V}.
The underlying mechanisms which drive a similar trend in the morphology-density relation in clusters may also exist in galaxy groups \citep[e.g.,][]{2003MNRAS.339L..29H, 2003MNRAS.340..485H, 1998ApJ...496...73M}.

	A wide variety of observations in both optical and X-ray regimes has provided convincing evidence that substructures in clusters are common. 
Rich galaxy clusters are believed to be built up by infalling galaxy groups \citep[e.g.,][]{1999ApJ...510L..15B, 1999A&A...343..733S, 1996ApJ...471..694A, 2005A&A...443...17A, 2006AJ....131..168K}, while galaxies in a group may have evolved already before the whole group falls into a cluster: i.e., what is referred to as \it pre-processing \rm \citep{2004PASJ...56...29F}. 
However, there is still considerable uncertainty on the roles that
these smaller structures play in the star formation history of cluster galaxies.
Therefore, cluster and group galaxy populations give us a unique opportunity to study the environmental effect on large (global, or `cluster') and small (local, or `group') scales.

In order to unfold the role of time and environment in galaxy evolution in clusters, we study these two major factors using both individual galaxies and galaxy groups in 16 clusters from the Canadian Network for Observational Cosmology (CNOC1) multi-band $BVR_cI$ photometric observations.
The structure of this paper is as follows. 
We present the data in $\S$\ref{dataset}. 
We then briefly describe the photometric redshift method and sample selection in $\S$\ref{sec:photoz}. The group finding method is briefly
presented in $\S$\ref{sec_pFoF}. The identification of cluster galaxies and groups is detailed in $\S$\ref{sec_group}. 
We present the calculations of the environmental parameters and red galaxy fraction in $\S$\ref{cnoc:fred}.
We describe the correction due to background contamination in $\S$\ref{background}.
We present the results $\S$\ref{results} and discuss their implications
in $\S$\ref{discussion}. Finally a summary is presented in $\S$\ref{conclusions}.
We adopt $\Omega_m$=0.7, $\Omega_{\Lambda}$=0.3, and $H_0$=70 km/s/Mpc.

\section{The Cluster Catalogs \label{dataset}}
\subsection{The Data}
  The original CNOC1 Cluster Redshift Survey \citep{1996ApJS..102..269Y} contains 16 rich X-ray selected galaxy clusters at $0.17<z<0.55$, and was carried out with the main goal of measuring the cosmological density parameter, $\Omega_m$. 
The observations were made using the CFHT Multi-Object-Spectrograph (MOS) with Gunn $g$ and $r$ images to $r\sim24$ and spectroscopy to $r\sim21.5$. 
The multi-color photometric follow-up observations in this study targeted 15 of the CNOC1 clusters using the CFHT 12K camera  in 2000 and 2001. 
The camera, composed of 12 chips of 2045 $\times$ 4096 pixels with 0.206$\arcsec$ per pixel, has a $42\arcmin\times 28\arcmin$ field of view. 
Ten clusters were taken from classical observations and the other five were from queue observations.
Among these 15 clusters, 14 were observed using $BVR_cI$ filters with seeing 
ranging from $\sim 0.70\arcsec$ to 1.24\arcsec~in the $R_c$-passband. 
MS0015+16 (CL0015+16, $z$=0.55) was observed in $B$ and $V$ only, and hence is excluded for not having enough passbands for our photometric redshift method. 
We also exclude the MS1224+20 field because the observations were non-photometric and not sufficiently deep for our scientific goals. 
MS1006+12 was omitted due to a very bright star in the field.
MS0302+16 ($z$=0.425, $\alpha_{2000}$=03:05:35.42, $\delta_{2000}$=17:10:3.43) was targeted incorrectly to MS0302.5+1717 ($z$=0.425, $\alpha_{2000}$=03:05:18.110, $\delta_{2000}$= 17:28:24.92), about 18.82\arcmin~north from the original CNOC1 target. 
Another cluster, CL0303+1706 (z=0.418, $\alpha_{2000}$=03:06:18.7, $\delta_{2000}$= +17:18:03), is in the south-east of MS0302.5+1717. 

 The pointings for MS1231+15 and MS1358+62 contain other clusters in the field. 
Abell 1560 \citep[$\alpha_{2000}$=12:34:07.1, $\delta_{2000}$=15:10:28, $z$=0.244;][]{1989ApJS...70....1A,1982ApJ...258..434U,1999ApJ...519..533D} is south of MS1231+15. MS1358+62 has been known as a binary cluster in the CNOC1 analysis \citep{1996ApJ...462...32C}. 
The other cluster in the MS1358+62 field is at $\alpha_{2000}$=13:59:39.511, $\delta_{2000}$=62:18:47.0, with  $z$=0.329. 
In summary, our data set contains 12 CNOC1 clusters and 4 other confirmed clusters. 
The sample properties and exposure times are listed in Tables \ref{tb1} and \ref{tb2}.

\subsection{Photometry and Color Calibration \label{cnoc:ppp}}
  For the queue-observed data, the basic data reduction and calibration steps
were undertaken using the \it Elixir \rm system at CFHT \footnote {http://www.cfht.hawaii.edu/Instruments/Elixir/realtime.html}.
For the classically observed data, the data reduction proceeded in the standard way using IRAF (http://iraf.noao.edu/) 
including bias subtraction and flat fielding. 
The photometric calibration was performed using images of standard star fields from \cite{1992AJ....104..340L}.

  The object finding, aperture photometry, and star-galaxy classification were carried out using the \it Picture Processing Package (PPP)\rm, details of which are described in \citet{1991PASP..103..396Y} and \citet{1996ApJS..102..269Y}. 
Here we present a brief summary of the procedures.
This algorithm detects objects by the locations of brightness peaks above a preset threshold level and satisfying a number of criteria (e.g., a minimum number of connected pixels, a sharpness test, etc., see Yee et al.~1996). 
In general, about 2,000 to 3,000 objects on average were detected in each chip. 
The photometry is achieved by constructing and analyzing the flux growth curve of an object. 
The total magnitude is measured in an `optimal aperture', and then corrected to a standard aperture of $8.5\arcsec$ diameter using the PSF profile of reference stars, if the optimal aperture of the object is smaller than the standard aperture. 
The color of each object is measured independently from the total magnitude, using a `color aperture' of $3\arcsec$, or the optimal aperture, whichever is smaller. 
The total magnitude for the second filter is then computed from the color and the total magnitude of the first filter. 
The star-galaxy classification is accomplished by comparing the growth curve of each object to a local set of bright, but not saturated, reference stars. 
The algorithm classifies objects into four categories: galaxies, stars, saturated stars, and spurious `non-object' (e.g., cosmic ray detections or cosmetic defects). 

  In order to check the photometry of the CNOC1 follow-up observations, we compare our $R_c$ data with the Gunn $r$ photometry obtained by the original CNOC1 program, using the transformation $r=R_c + 0.28 + 0.027\times(B-R_c)$ derived from stars in the M67 field. 
To check the color offsets in the other passbands, we use the calibrated $R_c$ photometry as the reference.
 For $B$ and $V$ photometry, we compare the star colors with those in the \it Red-Sequence Cluster Survey \rm \cite[RCS;][]{2005ApJS..157....1G} CFHT patches \citep{2005ApJS..158..161H} using bright stars ($18 \leq R_c \leq 22$). 
Because the RCS uses $BVR_cz'$ instead of $BVR_cI$ passbands, we compare the distributions of $I-R_c$ star colors with those in the MS1231+15, MS1455+22, and MS1512+36 fields. 
These three pointings have been processed using the \it Elixir \rm system, for which the $I$-band photometry is consistent with each other after being cross-checked. 
The zero points in $BVI$ are shifted if any color offsets are present. 

\subsection{Astrometry}
Astrometry for each frame is calibrated using the IRAF package \bf MSCRED \rm with USNO-a2 catalogs. 
The \bf MSCRED \rm package has the ability to operate on multi-chip CCD data simultaneously without the need to split each image frame into separate CCD images and process them individually. 
Therefore, each CNOC1 12K follow-up observation pointing is made into a multi-extension FITS with 12 extensions. 
An astrometry calibration file is generated by the \bf msctpeak \rm command based on the MS0302+16 field. 
After updating the image headers with the calibration file, the \bf mscczero \rm and \bf msccmatch \rm commands are used to apply the shifts and match the positions between the objects on the images and in the USNO-a2 catalogs. 
We also update the astrometry in the headers of the original CNOC1 MOS images, so the object list from the CFH-12K data can be easily matched with those spectroscopically observed using MOS.

\section{Photometric Redshift and Sample Selection \label{sec:photoz}}
  The photometric redshift technique allows us to estimate galaxy redshifts using photometry information. 
Because the information obtained from photometric redshifts is less accurate, the use of the technique requires careful consideration in selecting the samples. 
In this section we present briefly the photometric redshift technique and the sample selection; details can be found in Li \& Yee (2008). 

\subsection{Redshift Estimation}
  Galaxy redshifts are estimated using an empirical photometric redshift technique from Li \& Yee (2008). 
This technique assumes that redshift is a polynomial function of galaxy colors and magnitudes. 
The coefficients of this polynomial can be estimated by fitting the galaxies in a training set, which is a catalog containing galaxy spectroscopic redshifts and magnitudes.
Our training set contains 3988 galaxies in $BVR_cIz'$ up to $z\sim 1.4$ with $R_{c} < 24.$, constructed from the Hubble Deep Field \citep{2004ApJ...600L..93G} and CNOC2 catalogs \citep{2000ApJS..129..475Y}. 
The properties of the training set are described in \cite{2005ApJS..158..161H} and \cite{2008AJ....135..809L}. To achieve better-constrained solutions, the galaxies with spectroscopic redshifts in the CNOC1 sample (2022 galaxies in total) are added to the training set. 
The photometric redshift uncertainties and probability densities of individual galaxies, $P(z)$, are obtained as part of our photometric redshift technique as well.

\begin{figure}
        \includegraphics[width=12.7cm]{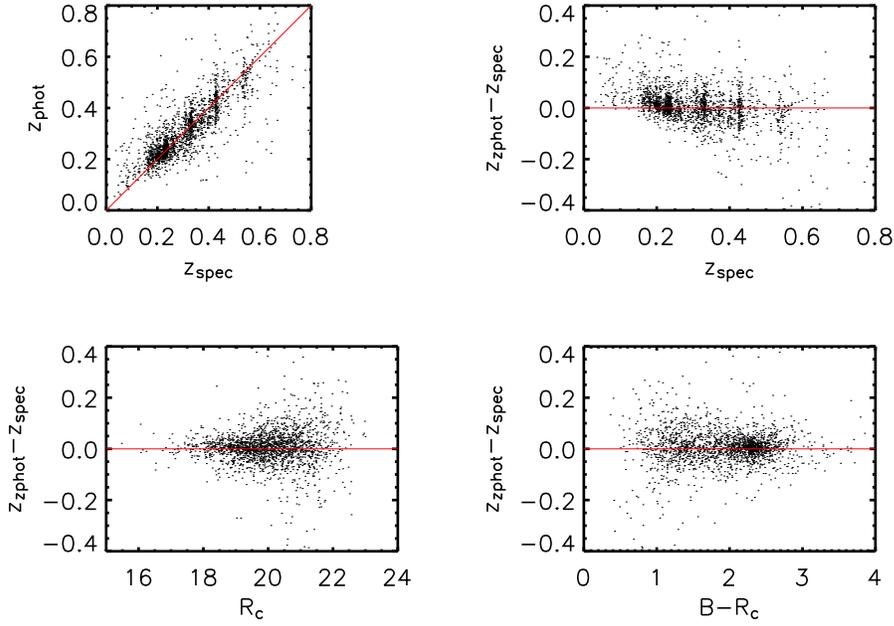}
        \caption[The photometric redshift measurement for the CNOC1 sample]{The estimated photometric redshift for 2022 CNOC1 galaxies with spectroscopic redshifts. The top two panels show the comparison between the spectroscopic and photometric redshifts. The overall dispersion is $\sigma(z_{phot}-z_{spec})\sim 0.051$. At $z<0.1$, the photometric redshift measurements are affected by not having a sufficiently blue band. The bottom two panels plot $z_{phot}-z_{spec}$ as functions of galaxy magnitude and color. The scatter is larger toward fainter magnitudes and blue colors: $\sigma (z_{phot}-z_{spec})\sim0.048$ and $\sim 0.108$ for $R_c < 21.5$ and $R_c \geq 21.5$, respectively, and $\sigma (z_{phot}-z_{spec}) \sim 0.041$ for $B-R_c \geq 1.8$ and $\sim 0.070$ for $B-R_c < 1.8$ \label{zphot.1}}.
	\end{figure}

  To assess the accuracy of our photometric redshifts, we compare them  with the spectroscopic redshifts of the 2022 CNOC1 galaxies.
The differences between photometric and spectroscopic redshifts are plotted in Figure \ref{zphot.1} as  functions of galaxy magnitude and $B-R_c$ color. 
The overall r.m.s., $\sigma(z_{phot}-z_{spec})$, is $\sim 0.051$, with $\sigma(z_{phot}-z_{spec}) \sim 0.048$ and $\sim 0.108$, for $R_c < 21.5$ and $R_c \geq 21.5$, respectively.
 Blue galaxies (defined by $B-R_c < 1.8$) has $\sigma(z_{phot}-z_{spec}) \sim 0.070$, while $\sigma(z_{phot}-z_{spec})$ is $\sim 0.041$ for red galaxies (defined
by $B-R_c \geq 1.8$). 

\subsection{Galaxy Sample Selection \label{subsec_sample}}
  For our analysis, we construct a sample of galaxies with a photometric
redshift range of $0.02 \leq z \leq 1.4$. 
The upper photometric redshift limit is due to the passband wavelength coverage for the $4000$ \AA~break in our training set. 
We select galaxies by their photometric redshift probability densities. 
The photometric redshift probability densities used are consistent with the observed scatter in $z_{phot}-z_{spec}$ in the training set for different types of galaxies. The selection criterion for galaxy $i$ is set as:
	\begin{equation}
   \int_{z_i - 3\sigma_{z}}^{z_i + 3\sigma_{z}} P_i(z') dz' > 0.997,
	\end{equation}
with $\sigma_{z}$ set to 0.2(1+$z$), which is equivalent to excluding galaxies whose photometric redshift uncertainty is greater than 0.2(1+$z$). 

Each galaxy in the sample is then assigned a completeness correction weight,  $w_i$, to account for galaxies excluded from the sample due to the $\sigma_{z}$ criterion.
The completeness correction is computed using the ratio of the total number
of galaxies to the number of galaxies 
 satisfying our selection criteria in magnitude bins of $\Delta m_{R_c}=0.1$. 
We set a cutoff in $R_c$ based on where $w_i=2$ to avoid high weight galaxies.
This apparent magnitude cutoff is brighter than the 100\% photometric completeness 
magnitude in all cases, 
and ranges from 22.12 to 22.80 for the different pointings.
The median and mean of the completeness correction factor $w_i$ in the 
sample are $\sim 1.08$ and $\sim 1.20$, respectively. 

The sample used for group finding is limited to galaxies with k- and evolution-corrected $R_c$-band absolute magnitude, $M_{R_c}^{k,e}$, brighter than $M^*_{R_c}+2$.
We adopt $M^*_{R_c}$=-21.41 at $z=0$ \citep{1997A&A...320...41K}.
The k-correction is computed using the tabulated values from \citet{1997A&AS..122..399P} for different spectral energy distributions (SED).
For each galaxy, we interpolate the k-correction values between the E/S0 and Sc SEDs based on its $B-R_c$ color.
We note that when selecting galaxies at different redshifts
using k-corrections derived from galaxy colors creates
a less biased sample than applying k-corrections averaged over SED types without
using color information \citep[e.g., see][]{2008ApJ...680..214L}.
The $R_c$ luminosity evolution is approximated as $M(z)=M(0)-zQ$, where Q=1.24 for red galaxies and Q=0.11 for blue galaxies \citep{1999ApJ...518..533L}.
In the cases when the apparent magnitude limit is brighter than $M^*_{Rc}+2$
at the photometric redshift of the group, we apply a correction 
factor $R_w$ which 
is the ratio of the integral of the galaxy luminosity function (LF) integrated
to $M^*_{Rc}+2$ to that integrated to the completeness limit 
(see Eqn 2 in Li \& Yee 2008), to correct for the expected missing galaxy counts 
in these fields.
Note that while we use $M^*_{R_c}+2.0$ as the limit of pFoF group finding
(which optimizes group-finding results), we limit the galaxy sample
to $M^*_{R_c}+1.5$ for our galaxy population analysis.

\section{Identifying Galaxy Groups \label{sec_pFoF}}
  Galaxy groups in the photometric redshift catalogs are identified using the \it `Probability Friends-of-Friends' \rm algorithm (pFoF); 
a modified friends-of-friends group-finding algorithm which takes into
 consideration the photometric redshift probability density of each galaxy and group. 
Detailed descriptions of the algorithm and performance tests
are presented in \citet{2008AJ....135..809L}; we provide a brief summary below.
The main idea behind this algorithm is to treat the common photometric redshift space of the transversely connected galaxies as the group redshift space, which changes dynamically as new members are linked in. 
This updated group redshift is then used to search for more new members. 
Generally speaking, the group redshift is well confined by the group members, especially when the number of its members is large. 
The linking criteria are set by two parameters: $D0_{xy}$ and $P_{ratio,crit}$. 
The $D0_{xy}$ is the reference physical transverse linking length at $z$=0, which is scaled by $(1+z)^{-1}$. 
In practice, the linking length includes the adjustments for the completeness correction, $w_i$, 
and the varying galaxy numbers at different redshifts due to the apparent magnitude cutoff. 
$P_{ratio,crit}$ is the criterion used in determining friendship in redshift space, where $P_{ratio}$ denotes the normalized total photometric redshift probability for a galaxy being in the same redshift space as another galaxy or a group
of galaxies (see Li \&~Yee 2008 for details).
The parameter $P_{ratio}$ essentially measures
the amount of overlap in the photometric redshift
probability distributions of the two objects being considered.
Galaxies must have $P_{ratio} \geq P_{ratio,crit}$ in order to join into the group.
We note that for the purpose of creating the group catalog
the algorithm assigns a group membership to every galaxy , with  isolated
galaxies being identified as groups of one.

This algorithm has been tested using mock catalogs constructed from the Virgo Consortium Millennium Simulation \citep{2005Natur.435..629S}, including the effects of using different linking criteria. 
With a fixed $P_{ratio,crit}$, the recovery rate increases when larger $D0_{xy}$ is used; but the use of a larger $D0_{xy}$ also increases the fraction of falsely detected groups. 
With a given $D0_{xy}$, the recovery rate is better when a smaller $P_{ratio,crit}$ is used, but the false detection rate is larger as well. 
As a compromise between the recovery and false detection rates, the set of $P_{ratio,crit}=0.37$ and $D0_{xy}=0.25$ Mpc gives the best results. 
With these parameters, the recovery rate is greater than 80\% for mock groups of halo mass $M_{halo}$ greater than $\sim 1.2\times 10^{13}M_{\sun}$, and the false detection rate is less than 10\% when a pFoF group contains $\sim 8$ 
or more net members (corresponding to $M_{halo}\sim 3.7\times 10^{13} M_{\sun}$).
The estimated pFoF group redshift uncertainty is $\Delta z_{grp}\sim 0.020$. 
We adopt $P_{ratio,crit}=0.37$ and $D0_{xy}=0.25$ Mpc for the group-finding procedure in this paper.

 For our analysis, we select groups with richness $N_{gal} \geq 8$ 
and $N_{gz} \geq 7$, where $N_{gal}$ is the net weighted galaxy number, and
$N_{gz}$ is the actual number of the linked group members returned by the pFoF algorithm. 
From the 13 pointings we identify a total of 188 groups 
satisfying our richness selection using the adopted pFoF parameters, 
including the cluster cores themselves.

\section{The Sample of Cluster Galaxies and Groups \label{sec_group}}
\subsection{The Cluster as the Main Group \label{subsec_CLmain}}
Generally speaking, galaxy clusters can be treated as exceptionally rich galaxy groups. 
Because our sample is selected from the follow-up observations of CNOC1 clusters with cluster redshift information available, we identify the galaxy group that coincides with the cluster center position as the main body, or core, of each cluster.

 The center of a group is determined using its members in the dense region where their local galaxy densities (see \S6.1)  are above the mean value of the group members. 
The center is calculated as the mean R.A. and Dec.~of these members, weighted by their luminosity, completeness correction $w_i$, and local galaxy density. 
If a group's members are distributed in more than one dense region (i.e., clumps in the contours of local galaxy density), only galaxies in the clump which has the largest galaxy number counts are used. 
To designate the main galaxy aggregation, or the core, of a cluster, each galaxy group is assigned a score based on: 
(1) the inverse of the separation between the group and cluster centers, (2) the agreement of the group redshift compared with the cluster spectroscopic redshift, and (3) the group richness.
The galaxy group which has the best score is chosen as
the cluster main-group, i.e., the core galaxy aggregation of a cluster.
We emphasize that cluster main-group associated with each galaxy
cluster does not contain the whole of the cluster, but rather just the
galaxies in the core that satisfy the linking criteria of the
pFoF algorithm.

The redshifts of our 16 main cluster galaxy aggregations ($z_{pFoF,cl}$) compared with the spectroscopic redshifts ($z_{spec,cl}$) have a dispersion of $\sigma(z_{pFoF,cl}-z_{spec,cl}) \sim 0.019$. 
The centers of our cluster main-groups are on average $\sim 30\arcsec$ from the physical centers (i.e., where the cD galaxies are). 
The basic properties of our 16 identified cluster main-groups are listed in Table \ref{Tmainbody}.

\subsection{Cluster Galaxies and Groups \label{subsec_grpselect}}
  To identify galaxies and groups in the cluster redshift space, we follow the same idea used in determining group membership in the pFoF algorithm (Li \& Yee 2008), but applied only in the redshift direction without considering the transverse friends-of-friends criterion. 
That is, the $P_{ratio}$ of each galaxy or group with respect to the 
redshift probability density of the cluster main-group should 
satisfy $P_{ratio} \geq P_{ratio,crit}$, with the same $P_{ratio,crit}$ value used in the pFoF algorithm. 
The galaxies and groups that satisfy this condition are then considered 
to be in the large scale structure that make up the cluster, 
including galaxies at large cluster-centric radii.

Applying the pFoF criteria to our group sample with respect to the cluster main-groups,
we find 89 groups (not counting the cluster main-groups) consistent
 with being in the same redshift space as the clusters.
Of these, 59 are within 3$R_{200}$ of the cluster centers. 
Similarly, we apply the pFoF criteria to individual galaxies brighter than $M^*_{Rc} + 1.5$ to create
a sample of galaxies considered to be in the redshift space of the clusters.
  Hereafter, by `cluster galaxies' or `cluster groups', we mean galaxies or groups whose photometric redshift probability densities within the cluster redshift space satisfy the $P_{ratio,crit}$ criterion.
Based on the overall group sample and the typical uncertainty
in the photometric redshifts for the groups, we estimate that
the cluster group sample may be contaminated by projection (i.e., groups
not associated with the same large scale structure in which
the the galaxy cluster resides) at a rate of $\sim0.25$ groups per 
cluster, sufficiently small that the effect can be ignored in our
analysis.

  We note that two pointings, 0302+16 and MS0451-03, are not complete 
to $M^*_{Rc}+1.5$ at the cluster redshifts.
For these clusters, we apply a correction factor, $R_w$ (see \S3.2), 
based on the galaxy LF, to correct the counts to the $M^*_{Rc}+1.5$ limit.
Note that this correction factor is not strictly correct
when used for estimating the red galaxy fraction \fred~(see \S6.3), 
since the red and blue cluster galaxy
LFs are not expected to be identical \citep[e.g.,][]{2007ApJ...671.1471B},
and will tend to decrease the estimated \fred~due to the steeper faint
end of the blue galaxy LF.
However, since our adopted absolute magnitude limit is relatively bright and 
the incompleteness ranges from 0.05 to 0.3 mag, this effect does not
significantly affect our conclusions.
 
\begin{figure}
        \includegraphics{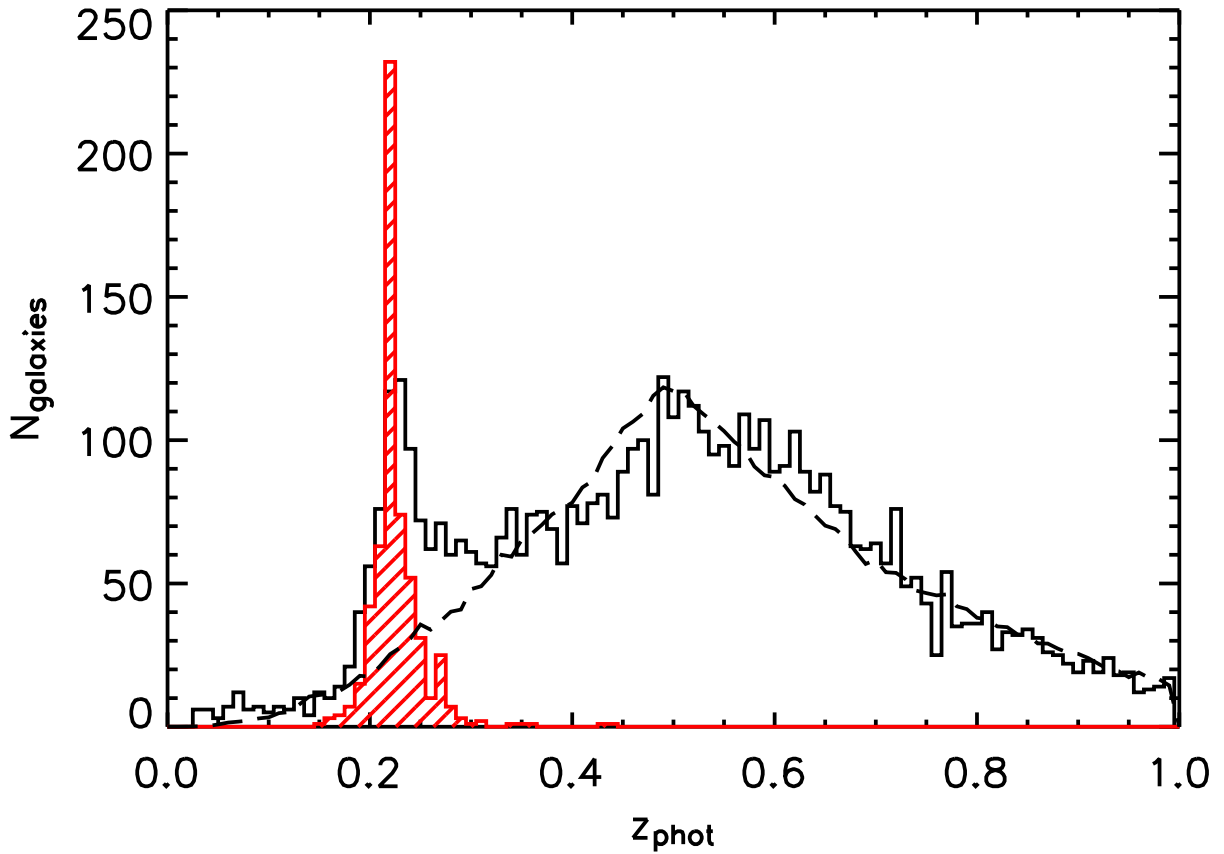}
        \caption[Photometric redshift distributions for galaxies in 
cluster Abell 2390]{The photometric redshift distribution (open histogram)
 of all galaxies with $m_{R_c} < 22.80$ in the field of Abell 2390 field ($z=0.23$). The background galaxy distribution estimated by the RCS CFHT patches (Hsieh et al. 2005) is shown as the dashed line. 
The cluster galaxies (hatched histogram), as determined by the group 
finding algorithm (see $\S$\ref{sec_pFoF}), show a large excess in
 galaxy number in photometric redshift space at the cluster redshift. \label{zpeak}}
        \end{figure}

  We present an example in Figure \ref{zpeak} using the cluster Abell 2390 ($z=0.228$) to illustrate the selection of cluster galaxies using photometric redshifts. 
The cluster galaxies, whose memberships are determined using the redshift assigned by the group finding algorithm, exhibit an excess peak in the photometric redshift space in contrast to the field galaxy distribution. 
Our method of selecting cluster galaxies and groups with reassignment of their redshifts to those of the associated pFoF groups eliminates foreground and background galaxies with higher efficiency
than the usual method of applying a photometric redshift cut based on the
individual galaxy's original photometric redshift and uncertainty.

\begin{figure}        
	\includegraphics{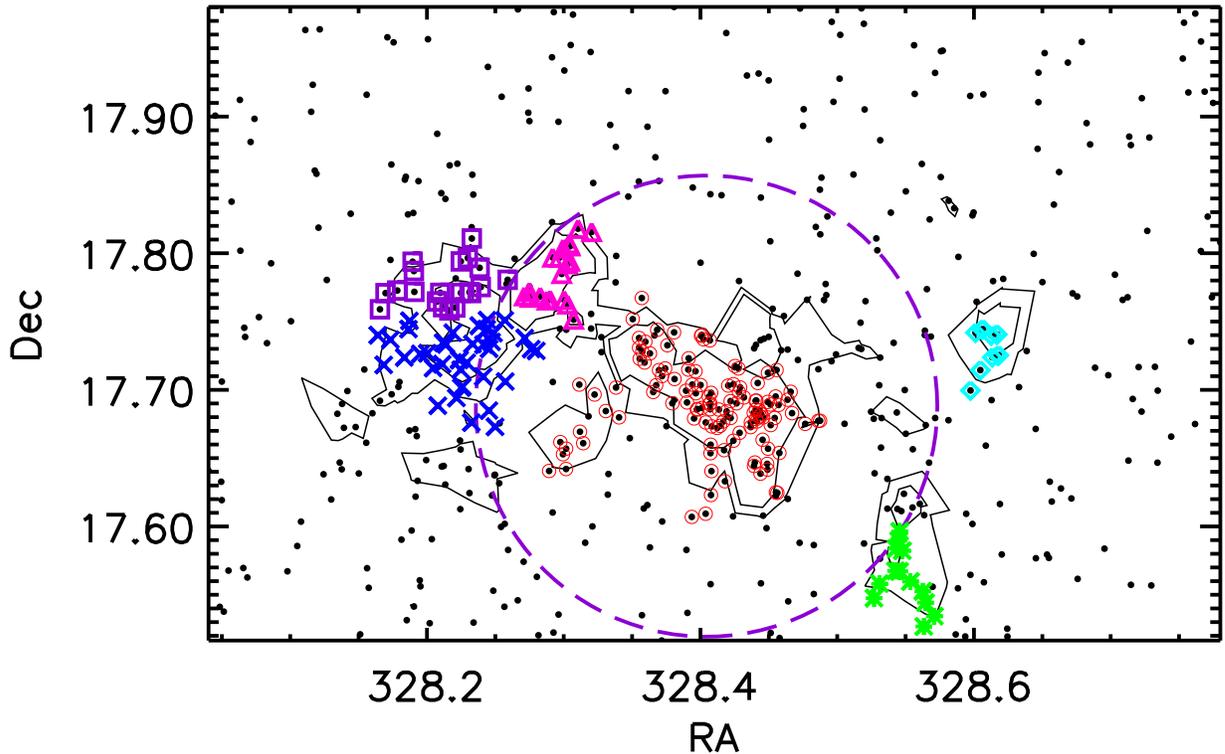}
        \caption[The map of Abell 2390]{The map of Abell 2390 with groups found using the pFoF algorithm. Galaxies selected in the cluster redshift space are plotted as dots.
The symbols with different shapes denote members of cluster groups 
with $N_{gal} \geq 8$. The cluster main-group members are marked by 
open circles. 
The contours of the local projected galaxy density $\Sigma_5$,  calculated using on all cluster galaxies, are overlaid with levels of $\Sigma_5$=20 and 40 gal/Mpc$^2$. The dashed circle has a radius of 1 $R_{200}$. \label{A2390.map}}
        \end{figure}

Figure \ref{A2390.map} shows the sky locations for the cluster galaxies and cluster groups in the Abell 2390 field. 
The cluster has an elongation from the central core at about --50$^0$ from the N-S axis, spanning both sides of the cD galaxy \citep{1996ApJ...471..694A,1996A&A...311..413P}. 
An infalling sub-component 
at the projected distance of $\sim 650\arcsec$ to the cluster center  
has been studied in \cite{1996ApJ...471..694A}.
This group is identified by the pFoF algorithm as well, and is marked by crosses in Figure \ref{A2390.map}. 
Our pFoF algorithm also identifies two additional groups in the same direction, which are marked as squares and triangles in Figure \ref{A2390.map}.
However, as they lie about 100\arcsec~and 280\arcsec~away from the spectroscopically identified group,  
these two groups do not contain any members in the original CNOC1 spectroscopic coverage. 

\begin{figure}
        \includegraphics{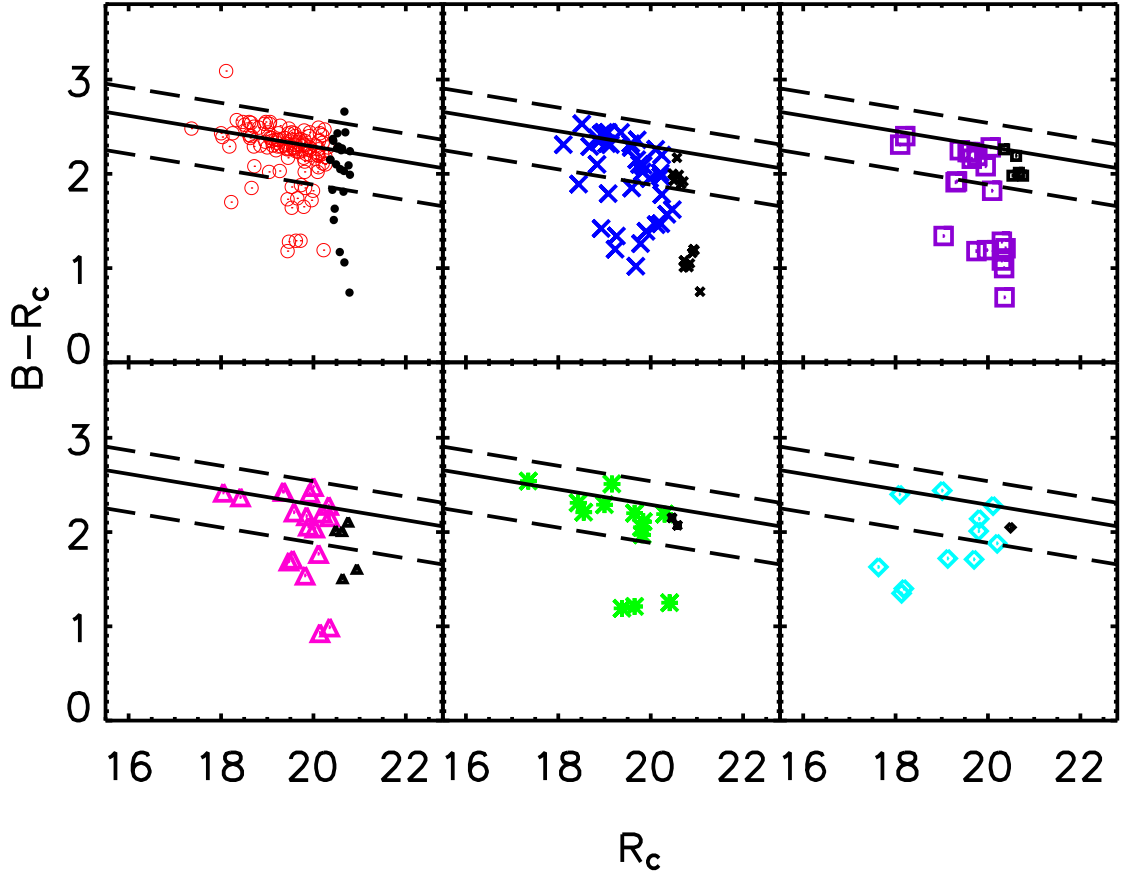}
        \caption{The observed color-magnitude diagram of galaxies ($M_{Rc}^{k,e} < M^*_{R_c}+2.0$) of galaxies in the cluster main-group and galaxy groups
in Abell 2390.
The theoretical red-sequence \citep{1997A&A...320...41K}, adjusted for 
calibration systematics (see \S6.3), is overlaid in each panel as the 
solid line, while the dashed lines define the region used for red galaxies
in the calculation of \fred. Group galaxies in each panel are plotted using the symbol corresponding to the same group in Figure \ref{A2390.map}.
Large symbols mark galaxies with $M_{Rc}^{k,e}<M^*_{R_c}+1.5$, which
are used for the computation of \fred.
	\label{A2390.map.fred}}
        \end{figure}

We also present the observed color-magnitude diagrams in Figure \ref{A2390.map.fred} 
for all these cluster groups.
Galaxies in the cluster main-group exhibit a well-defined red sequence, while galaxies in the other groups have visually somewhat less prominent red sequences, primarily
due to the small number of galaxies. 
We note that galaxies in groups in clusters may also include some 
galaxies from the main cluster in projection, which will produce systematically redder groups closer to the cluster core. However,  this effect on the average colors of the richer cluster groups used in our analysis (\S8.4) is likely to be minimal. 
The typical galaxy surface density for groups with 8 members or more
is 21.56 galaxies/Mpc$^2$, 
whereas the average non-group cluster galaxy surface density at $r_{CL}$=0.5-1.5
ranges from 2.43 to 5.95 galaxies/Mpc$^2$, and is as low as 2.05 
galaxies/Mpc$^2$ at the largest radii we probe.

\section{The Environmental Parameters and Red Fraction, \fred \label{cnoc:fred}}
   We intend to study the colors of cluster galaxies in different environments with the aim of identifying environmental effects on galaxy evolution. 
For this purpose, we demarcate the environmental effect into \it local \rm and \it global \rm regimes, and use the fraction of red galaxies as an indicator of their evolutionary stage. 
Here, we present our methods for computing the environmental parameters and the red galaxy fraction.

\subsection{Local Galaxy Environment: Local Galaxy Density $\Sigma_5$}
For each galaxy, the local galaxy density, $\Sigma_5$, is computed using a circular aperture and counting galaxies to $M_{Rc}^{k,e} < M^*_{Rc}+1.5$:
        \begin{equation}        
	\Sigma_5 = \frac{R_w \sum_{i}^{n} w_i}{\pi r_5^2} -  \Sigma_{bg}(z_{CL}). 
	\label{eq:sigma_5}        
	\end{equation}
In this calculation, \it n \rm is the $n^{th}$ nearest galaxy at a distance
$r_n$ from the seed galaxy such 
that the total completeness weight $\sum_{i}^{n} w_i \geq 5$.
When there is a non-unity weight correction, $r_5$ is not necessarily 
the distance $r_n$,
since the total number of objects at $r_n$ would not be the integer 5.
Assuming a uniform density, we can approximately
correct for the effect of the `missing galaxies' represented by the weight
corrections by computing the $r_5$ in Equation \ref{eq:sigma_5} as
	$r_5 = \sqrt{\frac{Rw \sum_{i}^{n} w_i}{n}} r_n$.
The computation of $\Sigma_5$ includes a background correction term
$\Sigma_{bg}(z_{CL})$ (see \S7).
In Figure \ref{A2390.map}, we also overlay the contours of local galaxy density to illustrate the substructures in the cluster redshift space for Abell 2390.

\subsection{Global Cluster Environment: Cluster-Centric Radius $r_{CL}$ \label{par_rcl}}
    The global cluster environmental influence is delineated using the cluster-centric radius parameter, denoted as $r_{CL}$, in units of $R_{200}$, where 
$R_{200}$ is the radius within which the mass density is 200 times the critical density. 
We use the $R_{200}$~values published in \citet{1997ApJ...485L..13C} for our CNOC1 clusters, except for MS0906+11. 
Note that these  are based on velocity dispersion measurements. 
The $R_{200}$ value for MS0906+11 is overestimated, due to the binary nature of the cluster, with two velocity structures appearing to be overlaid on the sky.
\citet{1999ApJ...527..561B} also analyze the velocity dispersions of CNOC1 clusters. 
In their work, they separate the two merging clusters in MS0906+11 
and present their respectively velocity dispersions as 886 and 725 km/s. 
Scaling with the value in \citet{1997ApJ...485L..13C}, we adopt $R_{200}$=1.79 Mpc for MS0906+11. 
For those clusters with unknown $R_{200}$, we estimate the values using a solution derived from the correlation between $N_{gal}$ and $R_{200}$: $N_{gal} \sim 11.8\times R_{200}^3$.
The $R_{200}$ values for the clusters are listed in Table \ref{Tmainbody}.

\subsection{Red Galaxy Fraction: $f_{red}$}
    In a given local-density environment or within a galaxy group, the red galaxy fraction, $f_{red}$, is computed using galaxies with $M_{R_c}^{k,e} < M^*_{R_c}+1.5$. 
We compute $f_{red}$ using the observed $B-R_c$ color-magnitude diagram.
First, an offset in the theoretical red-sequence zeropoint (from \citet{1997A&A...320...41K}) is applied based on our observed cluster data.
We then take the color half way between the SEDs of E/S0 and Sc as
the boundary separating the red and blue galaxies.
This boundary is close to the valley in the color distribution
between the red sequence and the blue cloud.
It ranges from  $\sim$0.39 mag at $z=0.2$ to $0.63$ mag at $z=0.5$ in $B-R_c$
to the blue of the red sequence.

Background galaxies play an important role in the estimation of the true $f_{red}$, especially in environments where the net  galaxy counts are small.
We therefore use Bayesian statistical inference based on Poisson statistics to estimate the $f_{red}$ probability function \citep{2004physics..12069D,2006MNRAS.365..915A}.
We then take the mean and the $68\%$ confidence interval of a $f_{red}$ probability function as the estimated $f_{red}$ value and its uncertainty. 

\section{The Control Sample \label{background}}
When photometric redshifts are used to identify cluster galaxies, the 
background galaxies still have a significant role in contaminating the sample. 
Such contamination, however, can be statistically estimated and corrected for using a large control sample. 
We use three RCS CFHT patches \citep[\it 0920\rm, \it 1417\rm, and \it 1614\rm; ][]{2005ApJS..157....1G, 2005ApJS..158..161H} as the control fields.
These three patches have the deepest photometry in the RCS survey and cover a total of 11.83 square degrees. 
To derive the galaxy surface density as a function of redshift, we first count the number of galaxies at each redshift, $N(z)$, within a bin size of 0.01 in $z$ by summing all the weighted photometric redshift probability densities within the redshift interval:
        \begin{equation}
        N(z) = \int^{z+0.005}_{z-0.005} P_i(z') w_i dz',
        \end{equation}                                                                       
where $P_i(z)$ is the photometric redshift probability density of each galaxy and $w_i$ is the complete weight (see \S3.2).
The surface density at each redshift $n(z)$ is accordingly obtained by dividing $N(z)$ by the total area of the RCS control samples at each redshift.
The $\Sigma_{bg}(z_{CL})$ in Eq. \ref{eq:sigma_5} is therefore computed as: 
	\begin{equation}
	\Sigma_{bg}(z_{CL}) = \int_0^{\infty} L(z_{CL})n(z_{CL}) dz_{CL} \label{Sigma_bg},
	\end{equation}
where $L(z_{CL})$ is the photometric redshift likelihood for which the members of a cluster main-group may occur (see Li \& Yee 2008).

When we conduct analyses using subsamples of cluster galaxies 
based on the local galaxy density, 
the background corrections are also derived using galaxies selected from the control sample with the same criteria.
For example, to estimate $N_{bg}$ within a $\Sigma_5$ region, the $n(z)$ in Eq. \ref{Sigma_bg} is computed using background galaxies with $\Sigma_5$ which are smaller than the upper boundary of the $\Sigma_5$ bin, 
instead of using \it only \rm background galaxies 
whose $\Sigma_5$ are within the same $\Sigma_5$ range.
This is because cluster galaxies in a high $\Sigma_5$ region may be contaminated by background galaxies from both high  and low-density regions, and those in low $\Sigma_5$ regions are only affected by background galaxies in low $\Sigma_5$ bins (otherwise their $\Sigma_5$ would be larger). 
The area of a $\Sigma_5$ clump, where $\Sigma_5$ is above a fixed cutoff, is computed using the same method in calculating group area in \citet{2008AJ....135..809L}. 

\section{Results \label{results}}
\subsection{The Butcher-Oemler Effect}
The Butcher-Oemler effect \citep{1984ApJ...285..426B} states that galaxy clusters at higher redshifts possess a larger fraction of blue members than those at lower redshifts. 
Instead of using the fraction of blue galaxies as the parameter to indicate 
the evolutionary stage of galaxies, we adopt the red galaxy fraction $f_{red}$, where a larger $f_{red}$ reflects a smaller blue galaxy fraction and vice versa. 
The Butcher-Oemler effect is suggested to have a dependence on cluster centric radius \citep[e.g.,][]{2001ApJ...547..609E,2004MNRAS.350..253D,2008ApJ...680..214L}, so we compute the $f_{red}$ for each cluster by using 
(1) members of each cluster main-group (as defined in \S5.1), 
(2) all cluster galaxies within $1 R_{200}$, and (3), within 1-1.5$R_{200}$.
We note that the cluster main-group members on average occupy an area 
smaller by a factor $\sim 4$ than that within $1R_{200}$; i.e., an area
with an equivalent radius of $\sim 0.5R_{200}$.
The $f_{red}$ are computed using galaxies brighter than $M^*_{R_c}+1.5$.
We note that this limit is identical to the original luminosity limit of $M^*_V=-20$ used by Butcher \& Oemler (1984) after correcting for the different $H_0$ used. 

\begin{figure}
        \includegraphics[width=12.7cm]{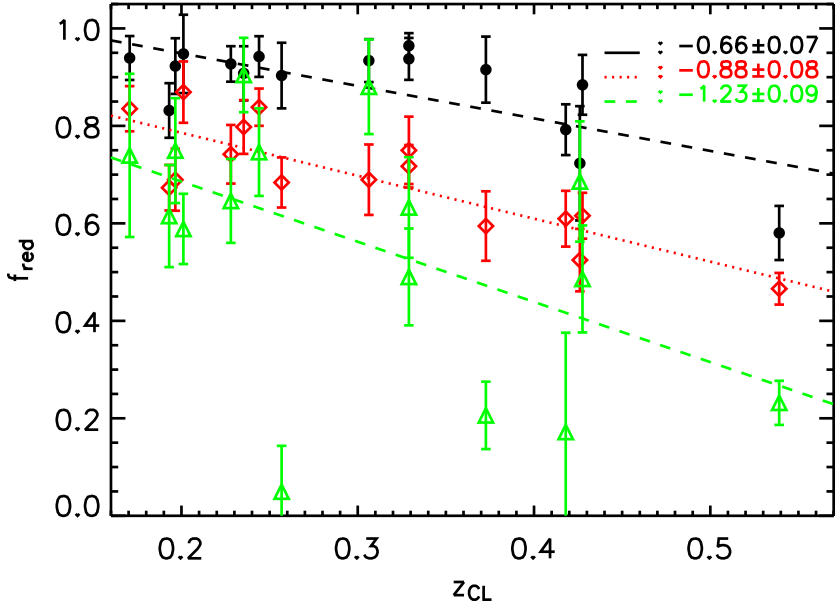}
        \caption[The $f_{red}$ in each CNOC1 cluster]{The red galaxy fractions for our 16 clusters using (1) galaxies in the identified cluster main-groups, displayed as solid circles; cluster galaxies (2) within 1 $R_{200}$, presented as  open diamonds; and (3) at 1-1.5 $R_{200}$, plotted as triangles. 
A linear fit between $f_{red}$ and cluster spectroscopic redshift is performed for each sample, with the slope and its uncertainty indicated on the plot.  
\label{BOplot}}
        \end{figure}

In Figure \ref{BOplot}~we plot the \fred~of each cluster versus its spectroscopic redshift for different cluster-centric radii.
We also overlay the results of linear fitting between $f_{red}$ and redshift. 
Figure \ref{BOplot} shows that there is a dependence of $f_{red}$ on the cluster-centric radius.
The $f_{red}$ computed using galaxies in the cluster main-group have 
the largest values, and this subsample also exhibits the most gentle gradient.
The $f_{red}$ for all three counting radii show a decreasing trend with redshift. 
It was shown in \citet{2001ApJ...547..609E} that the Butcher-Oemler effect 
is not significant within $0.5R_{200}$; our result of Figure \ref{BOplot} is consistent with their work.
For $z<0.40$, $f_{red}$ is essentially constant for the cluster main-groups 
(i.e., in the cluster core), but at higher redshift there appears to
 be a drop in the distribution of $f_{red}$, with two out 
of the four clusters at $z>0.40$ having $f_{red}$ significantly lower
 than the values from the low-redshift clusters.
This result can be interpreted as the $f_{red}$ in the cores of clusters reaching a `saturation' value close to 1 at $z\sim0.4$, while the Butcher-Oemler effect continues to be observed further out in radius. 

We note that the \fred~in the core of MS0451--03, at $z=0.539$, is a 
significant outlier, in that its value is comparable to those at larger radii.
This is likely a stochastic effect due to cosmic variance and not a
general redshift trend, as 
the other CNOC1 cluster at the same redshift, MS0016+16 (for which we 
do not have photometric redshift data), 
shows a considerably higher red galaxy fraction in the core, derived using the CNOC1 spectroscopic data,
than that of MS0451-03 \citep[see][]{2001ApJ...547..609E}.
A sudden change in the average core red fractions of clusters
is not seen in the large sample of clusters at $0.45<z<0.90$ by
\citet{2008ApJ...680..214L}, further suggesting that MS0451--03 is
significantly bluer in the core than the average for clusters at this redshift.

\subsection{The Butcher-Oemler Effect and Stellar-Mass Selected Samples}
The Butcher-Oemler effect has brought much debate in the literature since its first study in 1984 \citep[e.g.,][]{2001ApJ...548L.143M,2000AJ....119.1090M,2004MNRAS.351..125D,2005ApJ...619..134T,2008ApJ...680..214L,2005ApJ...626L..77N}. 
The most controversial one is its existence: whether the Butcher-Oemler effect is a real phenomenon or simply due to sample selection effects \citep[e.g.,][]{1999ApJ...516..647A, 2004MNRAS.349..889A}. 
For example, there are studies showing that the Butcher-Oemler effect becomes insignificant when the galaxy sample is selected using K-band luminosity \citep[e.g.,][]{2003ApJ...598...20D}.
More recently, \cite{2007ApJ...670..190H}, using a small sample of 5 rich 
clusters with $0.02<z<0.9$ and sampling with a relatively high 
stellar-mass limit of 10$^{10.6} M_{\sun}$,  
concluded that while the Butcher-Oemler effect is seen in a luminosity-selected cluster galaxy sample, a stellar-mass selected sample show insignificant changes in the galaxy red fraction over the redshift range.

\begin{figure}
	\includegraphics{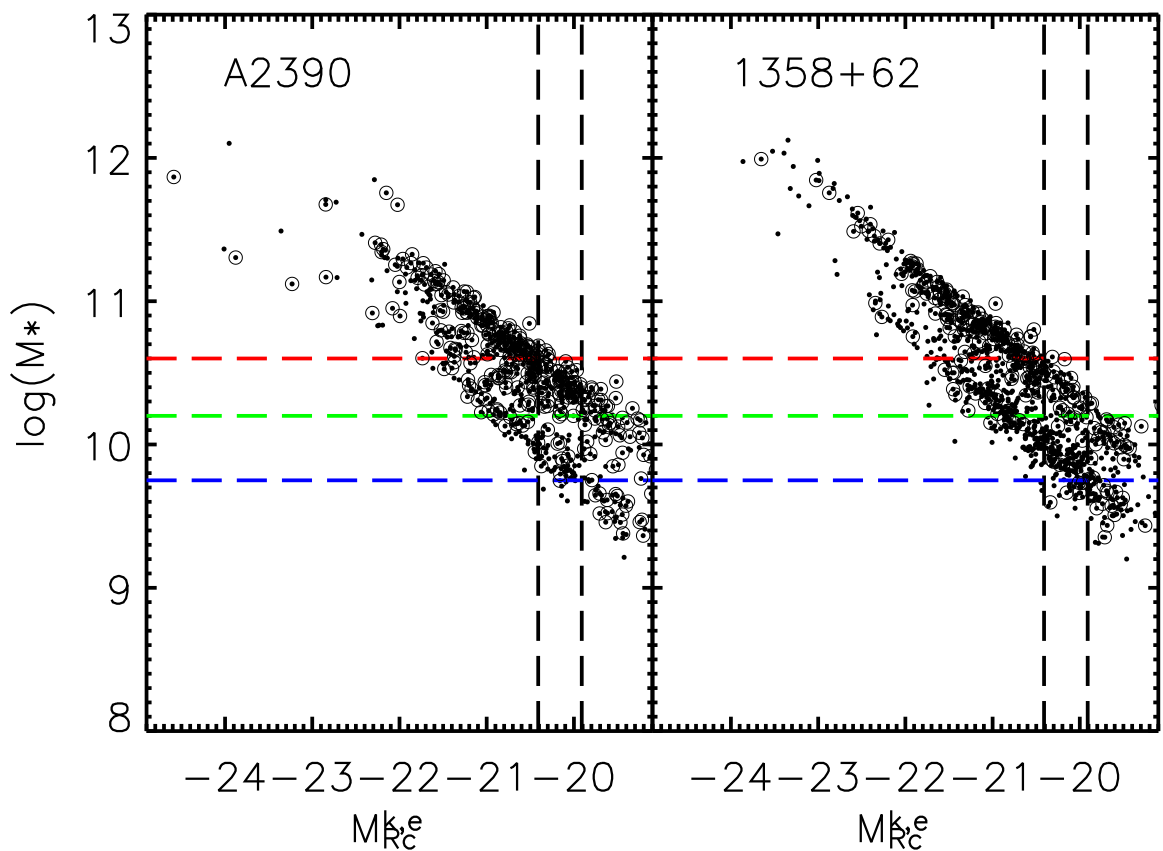}
	\caption{The estimated stellar mass M$_*$ as a 
function of $M_{Rc}^{k,e}$ for Abell 2390 and MS1358+62. The dots are all cluster galaxies in the patch and the open circles mark those within 1$R_{200}$ of the cluster center. 
The vertical dashed lines indicate $M_{Rc}^*+1.0$ and $M_{Rc}^*+1.5$. 
The horizontal dashed lines mark log M$_*$ = 9.75, 10.2, and 10.6, respectively, increasing from the bottom.
	\label{massR}}
	\end{figure}

To examine this issue, we investigate the effect of stellar-mass selection on the Butcher-Oemler effect for the CNOC1 sample.
 We derive the stellar mass in solar units, M$_*$, for galaxies in the cluster galaxy sample
by applying the algorithm in \cite{2003ApJS..149..289B}.
\cite{2007ApJ...670..190H} used the same method, and provided a detailed
discussion of the robustness and uncertainties in the derived stellar mass.
We use the relation: log$({\rm M}_*/L_{Rc})$ = --0.523 + 0.686($M_B$ - $M_{Rc}$) from \cite{2003ApJS..149..289B} 
with the values of the solar units from \citet{1994ApJS...95..107W}.
Both $M_{Rc}$ and $M_B-M_{Rc}$ in our computation
are k-corrected using color-dependent k-corrections 
(see $\S$\ref{subsec_sample}).
We note that using the redder $Rc$-band luminosity, compared to the
$B$ band which was used by \cite{2007ApJ...670..190H}, produces more robust mass
estimates with less scatter.
As examples, we present in Figure \ref{massR} the relationship between $M_{Rc}^{k,e}$ 
and stellar mass for Abell 2390 and  MS1358+62.
In the Figure we also show lines marking the $M_{Rc}^{k,e}$ sampling limits of
$M^*_{R_c}+1.0$ and $M^*_{R_c}+1.5$, and stellar mass cuts at log M$_*$ = 
10.6, 10.2, and 9.75.
We note that the mass cuts at 10$^{10.2}$ and 10$^{9.75}$ $M_{\sun}$~correspond approximately to the average stellar mass for red and blue galaxies at $\sim M^*_{R_c}+1.5$, respectively;
whereas a stellar mass of 10$^{10.6}$$M_{\sun}$ coincides with red galaxies at $\sim M^*_{R_c}+1.0$.
Other clusters have remarkably similar plots, with the major difference being the completeness level (with most lower redshift clusters being considerably
more complete than  $M^*_{R_c}+1.5$).
The most incomplete cluster is MS0451-03, which is complete to about $M^*_{Rc}+1.2$.  We note that our $w_i$ and $R_w$ correction factors partially alleviate
this problem.

\begin{figure}
	\includegraphics{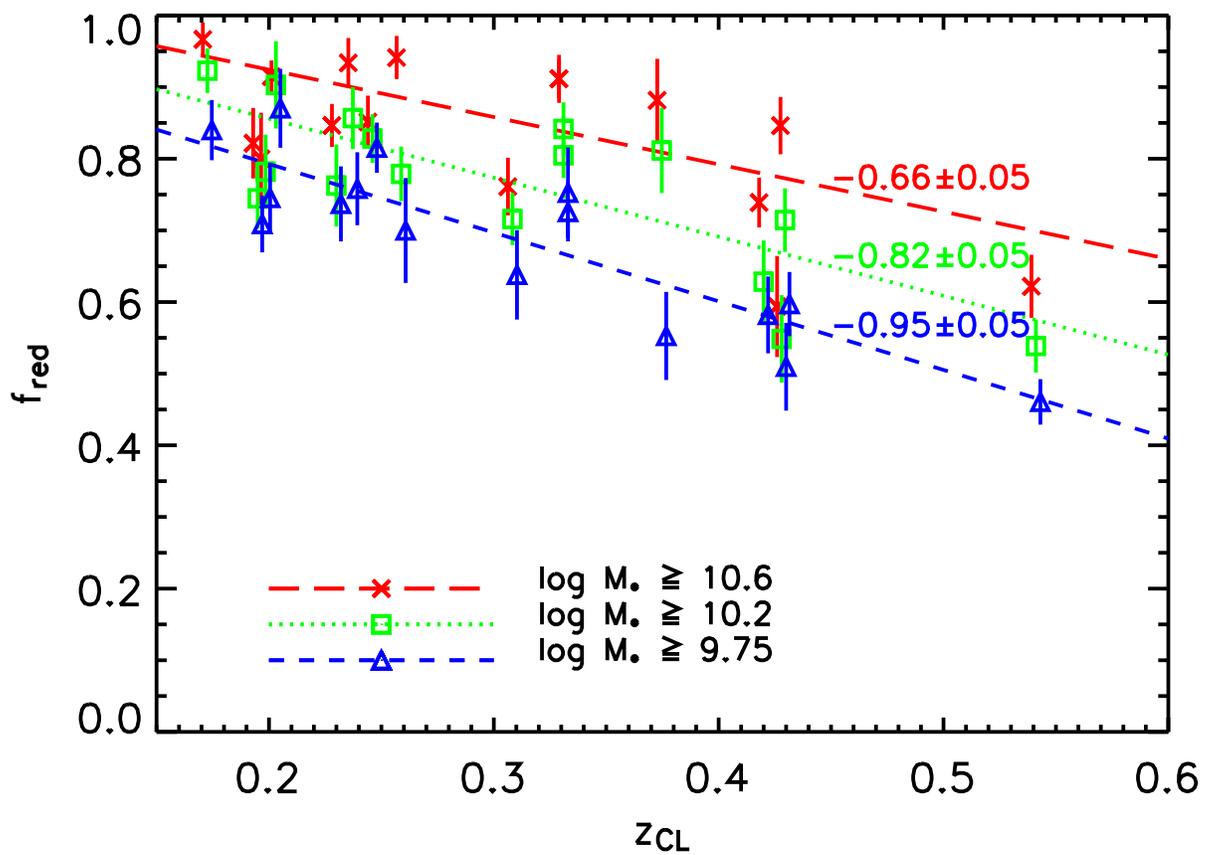}
	\caption{The $f_{red}$ for galaxies within 1$R_{200}$ of each cluster
 as a function of redshift. Results using  three different
stellar-mass (M$_*$) limits are shown. 
A linear fit between $f_{red}$ and cluster spectroscopic redshift is performed 
for each sample, with the slope and its uncertainty indicated on the plot.  
	\label{BOmstellar}}
	\end{figure}

Using the stellar-mass selected sample for galaxies within 1$R_{200}$, we re-derive \fred~using a statistical background subtraction method similar to that
for the luminosity-selected sample.  
Note that because of the considerably larger uncertainties in the photometric redshifts of the control sample galaxies, the derived stellar-mass measurements are expected to have significantly larger scatter.
We show the results in Figure \ref{BOmstellar}, where we plot the red galaxy fractions for the three different stellar-mass cuts.
All three samples show the Butcher-Oemler effect, with the lower mass limit sample showing the largest change in \fred~with redshift.
We note that there is significant incompleteness in the red galaxies for the
higher redshift clusters for the lowest mass cut (log$M_*$ = 9.75) where our incompleteness correction procedure may not be effective and hence may produce smaller \fred~for the higher-redshift clusters.
However, the  log M$_*$ = 10.2 sample, which is largely complete, also shows that a lower mass cut produces a more significant Butcher-Oemler effect.
The result with the log M$_*$ $\geq$ 10.2 sample is essentially identical
to that in Figure \ref{BOplot} (open diamonds, using galaxies inside 1$R_{200}$
with a luminosity limit of $\sim M^*_{Rc}+1.5$), except for a systematic
 shift to slightly higher average \fred~values, 
which is due to the larger number of red galaxies in the sample
when a stellar-mass cut is used.

Our sample shows that the slope of the \fred-$z$ trend is significantly
shallower when using a higher stellar-mass limit.
This is not a surprising conclusion, and is expected from the
well-established observations pointing to the down-sizing scenario of galaxy evolution 
\citep[e.g.,][]{1996AJ....112..839C,2006MNRAS.366..499D,2004MNRAS.350.1005K,2003ApJ...597..730L} and 
the build-up of the faint end of the red sequence with decreasing redshift
\citep[e.g.,][]{2008ApJ...673..742G,2008MNRAS.387...79V,2007MNRAS.374..809D,2005MNRAS.362..268T,2004ApJ...610L..77D}.
 Both of these phenomena
would naturally reduce the observed Butcher-Oemler effect,
if the chosen sample has too high a limit in stellar mass or luminosity.
Hence, the lack of strong evolution in the \cite{2007ApJ...670..190H} result may arise from their relatively high stellar-mass limit.
The small sample size of 5 clusters used by Holden et al.~also makes
detecting a reduced Butcher-Oemler effect more difficult.
Furthermore, the Holden et al.~results are based on cluster galaxies within
a cluster-centric radius of 1.25Mpc, likely significantly smaller than
the $R_{200}$ radii of these rich clusters.
This will also further reduce the detectability of the Butcher-Oemler effect.

\begin{figure}
	\includegraphics{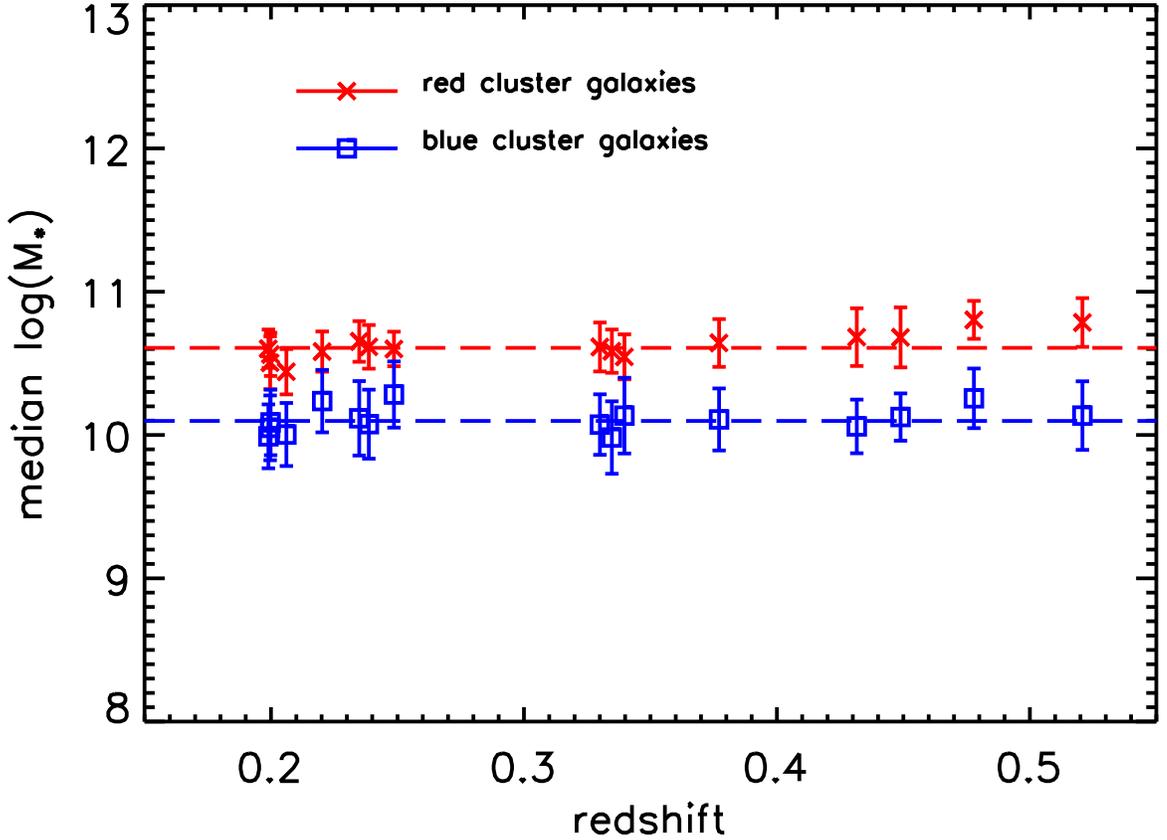}
	\caption{The median stellar mass as a function of redshift for
red and blue galaxies in the clusters.
Galaxies are selected within 1$R_{200}$ and in the absolute magnitude range
$M^*_{Rc}+0.5\le M^{k,e}_{Rc}\le M^*_{Rc}+1.5$.
Red galaxy samples are represented by crosses, and blue galaxy samples by
open squares.
These median stellar-mass values provide relative estimates of the
stellar-mass limits for the red and blue galaxies 
in the cluster sample at the luminosity limit of $ M^*_{Rc}+1.5$.
The horizontal dashed lines show the median of the values from all the clusters.
 The stellar-mass limits are similar for all
the clusters in the sample.  The small increases seen in the limits for the
highest redshift clusters are due to slight incompleteness in absolute
magnitude.
	\label{massz}}
	\end{figure}

Figure \ref{massR} also demonstrates that a sample chosen using a SED-dependent
k- and evolution-corrected luminosity limit can in fact provide an unbiased 
measurement of the Butcher-Oemler effect, if this effect is generalized to mean the evolution of the color fraction of galaxies.
For a chosen $M_{Rc}^{k,e}$ limit, there would be different limits for the average stellar mass for the red sequence galaxies and for the blue galaxies.  
As long as these mass limits do not change as a function of redshift, 
an $M_{Rc}^{k,e}$-selected sample essentially samples similar galaxies over the redshift range, and hence the evolution of \fred~is not driven by selection effects.
As an example, for the CNOC1 sample presented here, 
our k- and evolution-corrected $M_{Rc}^{k,e}$ limit samples red and blue
galaxies to stellar-mass limits of log M$_*$ of $\sim$
10.2 and 9.75, respectively, for all our clusters.
Figure \ref{massz} plots the median stellar mass in the magnitude bin
$M^*_{R_c}+0.5<M_{Rc}^{k,e}<M^*_{R_c}+1.5$ for the red and blue cluster galaxy samples as a function of cluster redshift.
This illustrates that there is no substantial selection effect in the stellar mass sampled as a function of redshift.
The somewhat higher mass value for MS0451-03 is due to incompleteness 
in $M_{Rc}^{k,e}$ in the highest redshift cluster.

In the remainder of the paper, we analyze the cluster galaxy stellar population using the luminosity-selected sample defined in $\S$\ref{subsec_sample},
which is equivalent to sampling mass limits of  log M$_*$ = 10.2, and 9.75
for red and blue galaxies, respectively. 
This allows for a more robust background subtraction based on
observed magnitudes, and also increases the number of galaxies used 
in deriving the various statistics.

\subsection{Cluster Galaxies \label{res_CLg}}
In this subsection, we examine the dependence of galaxy populations on
redshift, $r_{CL}$, and $\Sigma_5$ using samples of galaxies combined
from the different clusters.
The cluster galaxies are separated into three redshift bins:
 $0.15 < z < 0.30$, $0.30 \leq z < 0.40$, and $0.40 \leq z < 0.55$, 
with 8, 4, and 4 clusters in each bin, respectively. 
 
\begin{figure}
	\includegraphics{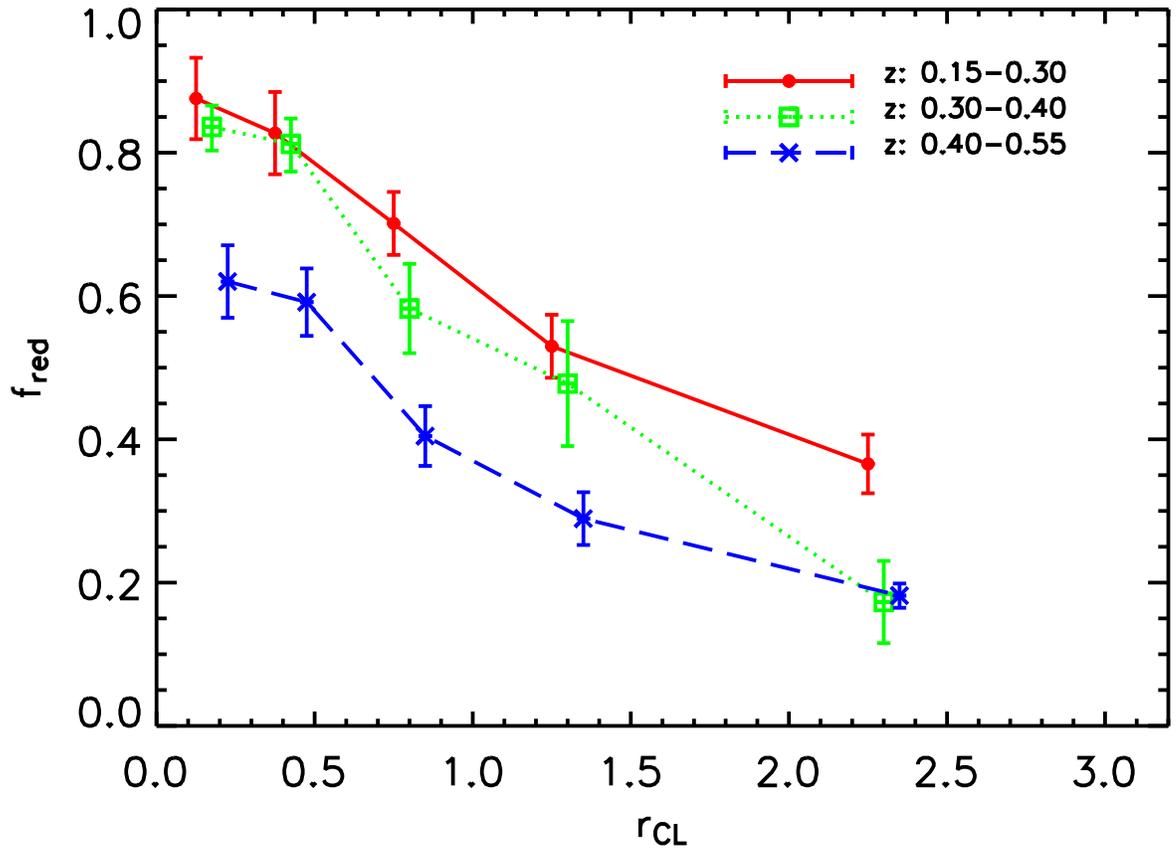}
	\caption{The $f_{red}$ as a function of $r_{CL}$ in three redshift bins, as indicated on the plot. The $r_{CL}$ values in each redshift bin are shifted slightly to provide clarity.
Galaxies in all $\Sigma_5$ regions are included.  \label{rfred}}
	\end{figure}

\begin{figure}
	\includegraphics{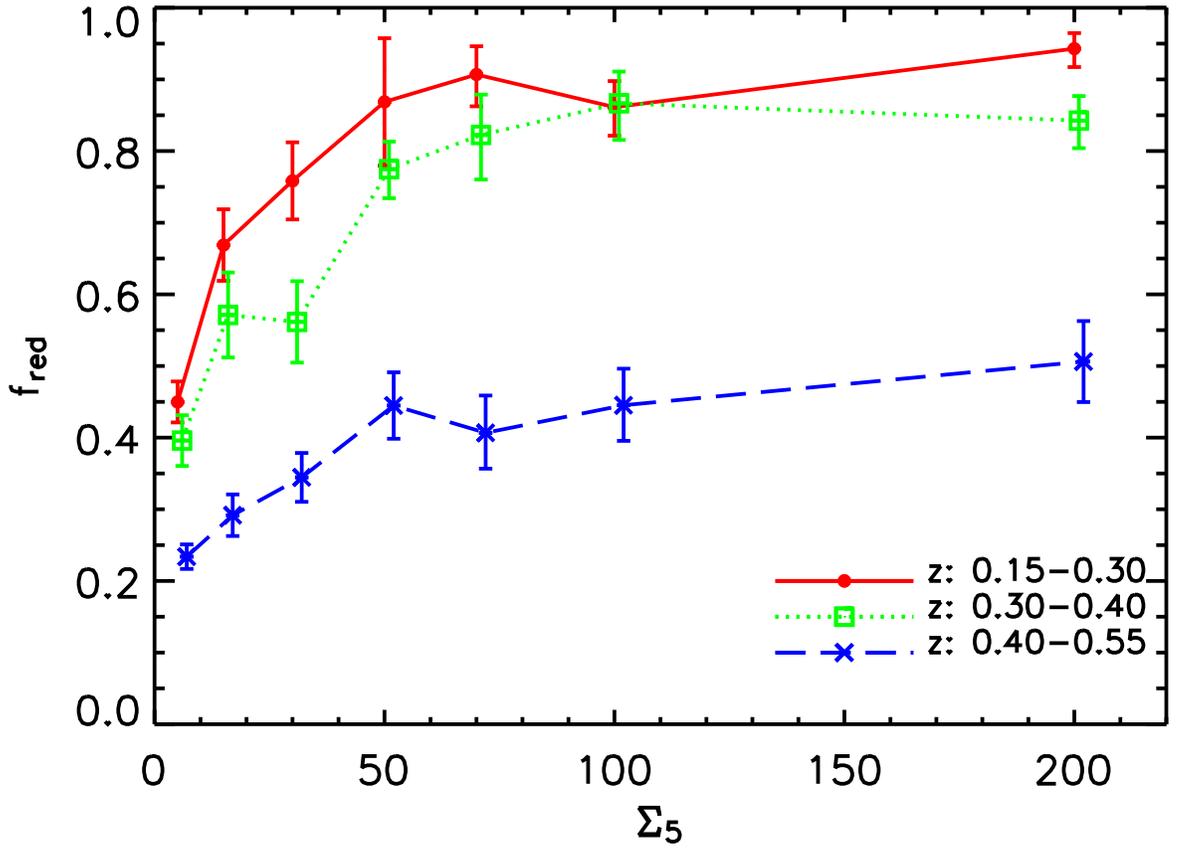}
	\caption{The $f_{red}$ as a function of $\Sigma_5$ in three redshift bins, as indicated on the plot. The $f_{red}$ increases with increasing $\Sigma_5$ in all three redshift bins. \label{dfred}}
	\end{figure}

We plot in Figure \ref{rfred} the radial dependence of $f_{red}$ at each redshift bin using galaxies in all local density environments. 
As expected, there is a strong radial dependence of \fred~on cluster-centric radius for all redshift bins.
Furthermore, we see a Butcher-Oemler effect at all radii between the high- and low-redshift bins.
Figure \ref{dfred} shows $f_{red}$ as a function of $\Sigma_5$ in each redshift bin using all cluster galaxies within $r_{CL}\le 3 R_{200}$.
The galaxy red fraction in general  increases with $\Sigma_5$.
This is consistent with the well-established observational trend that 
that early-type, non-star-forming
  galaxies tend to populate high $\Sigma_5$ regions, while late-type,
 star-forming galaxies are more common in low $\Sigma_5$
 regions \citep[e.g.,][]{1980ApJS...42..565D, 2001ApJ...562L...9K, 2007MNRAS.tmp..224C, 2007ApJ...670..190H,2003ApJ...584..210G, 2005ApJ...629..143B}.
Figure \ref{dfred} also shows that the magnitude of the Butcher-Oemler effect depends on the local galaxy density.
The high-density regions show a stronger increase in $f_{red}$ as 
a function of redshift, with the highest $\Sigma_5$ bins reaching 
the nominal `saturation' $f_{red}$ values close to 1 at $z\sim0.40$; 
whereas the lower density bins show a progressive increase in \fred~over 
 the whole redshift range.
The trend in Figure \ref{dfred} can be interpreted as galaxies in 
higher and higher  density regions completing their star formation
 (or have their star formation quenched) at earlier and earlier epochs.  
For example, galaxies with $\Sigma_5\ge100$ reach $f_{red}\sim0.9$ by $z\sim0.35$, 
while those with $50\le\Sigma_5<100$ become mostly red the lower redshift of
 $\sim0.20$.

Because the apparent effects of  $r_{CL}$ and $\Sigma_5$ are largely degenerate,
we probe the change in $f_{red}$ in different environments in more detail 
by computing its values in bins of $\Sigma_5$ within a fixed $r_{CL}$ location, and  as a function of redshift.
We divide the data into bins with $r_{CL} < 0.50$, $0.50 \leq r_{CL} < 1.0$, $1.0 \leq r_{CL} < 1.5$, and $1.5 \leq r_{CL} < 3.0$.
The data are also separated into bins of $0 \leq \Sigma_5 < 10$, 
$10 \leq \Sigma_5 < 20$, $20 \leq \Sigma_5 < 40$, and $\Sigma_5\ge 40$.
Figure \ref{zplot} presents the \it `$f_{red}$-z' \rm trend in each $r_{CL}$ and $\Sigma_5$ bin. 
In all but one panel, $f_{red}$ declines with increasing redshift, indicating that within the uncertainties of the data the Butcher-Oemler effect is observed over the whole ranges of $r_{CL}$ and $\Sigma_5$ values.
At large cluster-centric radii, we see that the Butcher-Oemler effect becomes
progressively steeper with increasing local galaxy density.
In contrast, in the cluster core the galaxy populations are 
dominated by red galaxies over most redshift bins, 
and  show little evolution for $z< 0.35$ for all local densities.

We examine the effect of local galaxy density on $f_{red}$ further 
in Figure \ref{dplot}, where we plot the $f_{red}$ as a function of $\Sigma_5$ (the \it `$f_{red}$-$\Sigma_5$' \rm trend) in different $r_{CL}$ and redshift bins. 
In general we find an increase of $f_{red}$ with increasing $\Sigma_5$, as expected from previous studies.
However, the magnitude of the effect appears to be weak, and dependent on both the cluster-centric radius and the redshift.
At each redshift, the slope of the  `$f_{red}$-$\Sigma_5$'~trend is steepest at the outskirts ($1.5<r_{CL}<3.0$), and very weak or non-existent for galaxies within the virial radius of the clusters.
An interesting observation is that the overall increase in $f_{red}$, as one moves from the outer radial bins to the inner ones, is greater than, or comparable to, the increase in $f_{red}$ as a function of $\Sigma_5$ over the entire local galaxy density range.
This, along with the general lack of a strong dependence of $f_{red}$ 
on $\Sigma_5$ in the inner parts of the clusters, implies that the 
influence of the global cluster environmental, rather than the local galaxy 
density effect, dominates inside the cluster virialized region. 
Compared with the strong \it `$f_{red}$-$\Sigma_5$' \rm~trends shown
in Figure \ref{dfred}, the weak correlations of \fred~with $\Sigma_5$ seen
when the data are divided into $r_{CL}$ bins (Figure \ref{dplot}) 
suggest that at least part of the $f_{red}$-$\Sigma_5$ dependence in Figure \ref{dfred}
is the result of a cluster-centric radius effect, 
since galaxies in high-density regions have a higher chance of 
being found in small $r_{CL}$ locations.

In Figure \ref{rplot}, we plot $f_{red}$ as a function of $r_{CL}$ 
in different redshift bins to examine more explicitly the effect of the
 global cluster environment on galaxies, with the local galaxy density 
controlled.
We find that, in general, $f_{red}$ decreases with increasing $r_{CL}$ for
all local densities.
For the two lower-redshift samples,
high $\Sigma_5$ regions have, on average, flatter \it `$f_{red}$-${r_{CL}}$' \rm trends than those for the low local density bins. 
However, the \it `$f_{red}$-${r_{CL}}$' \rm trends for the high-redshift
sample are relatively steep for all the density bins.
This indicates that, over our redshift range, the apparent effect of the 
global cluster environment is strongest in low local-density regions.
This result suggests that galaxies in high local-density regions (e.g., galaxy groups) are likely to be already in a more advanced evolutionary stage, 
i.e., with large $f_{red}$, 
before infalling into the cluster, and hence the global
 effect of the cluster on their $f_{red}$ would be less apparent.

\subsection{Cluster Galaxies in Galaxy Groups \label{CL_Ggrp}}
From the results of \S\ref{res_CLg}, it is apparent that cluster galaxies in high local density regions have a different evolutionary history from those in low-density environments; however, there is also evidence that the observed galaxy density may not be the primary parameter producing these effects.
High galaxy density regions outside the cluster core are often indications of galaxy groups; therefore, it is useful to attempt to delineate the 
effects of a group environment from those of a high local galaxy density environment.

     To probe the role of galaxy groups in the evolution of cluster galaxies, we flag cluster galaxies into `group' and `non-group' cluster galaxies according to the pFoF group finding results (see $\S$ \ref{subsec_grpselect}).
Group cluster galaxies are the cluster galaxies classified as the members of pFoF groups of $N_{gal} \geq 8$ and $N_{gz} \geq 7$, 
and non-group cluster galaxies are those considered by the group-finding 
algorithm to be in pFoF groups with  $N_{gal} \le 2$ and $N_{gz} \le 2$.
Recall that the pFoF algorithm assigns a ``group" membership to all
galaxies, including isolated galaxies (which are assigned to groups
with one member).
The gaps in the $N_{gal}$ and $N_{gz}$ cutoffs serve to reduce the ambiguity in the group membership of the galaxies. 
The relatively high $N_{gal}$ criterion, which corresponds to
a halo mass of $\sim3.7\times 10^{13}$\msun, is chosen so that the
group sample has an acceptable false detection rate of $<10$\%.
The low $N_{gal}$ criterion ensures that the non-group cluster galaxies
are relatively isolated and in subhalos with no more than two
galaxies brighter than $M^*_{R_c}+2.0$.
Scaling with $N_{gal}$, we expected these galaxies to be in halos
with mass less than $10^{13}$\msun.
The pFoF groups with 3 to 7 members (the `in-between' sample) are less well
defined, as they 
 have high false detection rates, ranging from 70\% to 20\%.
These groups are expected to have a broad halo mass distribution,
from that of a single galaxy to $\sim3\times 10^{13}$\msun.
The cluster galaxies associated with the cluster main-groups are excluded from the group cluster galaxies, so that the environmental effects caused by galaxy groups can be specifically investigated. 
We can consider that cluster main-group galaxies are the galaxies residing
in the core of the main dark matter halo which defines the cluster.

We have a total of 1073, 1226 and 3507 cluster galaxies in the cluster main-group, group, and non-group subsamples, respectively.
A total of 5727 galaxies fall in between the group and non-group criteria.
It is interesting to note that with this subdivision of cluster galaxies,
we find that $\sim 17.5$\% of the galaxies outside the virial radius
($> 1 R_{200}$) are in subhalos with masses 
greater $\sim 3$--$4\times 10^{13}$\msun.
This is broadly consistent with the simulations of
\citet{2009ApJ...690.1292B} which found $\sim$15\% of 
galaxies in their most massive clusters (which are most comparable
with the CNOC1 sample) are accreted from 
groups with halo masses larger than $\sim 3.5\times 10^{13}$\msun .

We illustrate in Figure \ref{A2390.den} 
an example of the distributions of the three subsamples of cluster
 galaxies in the $\Sigma_5$-$r_{CL}$ plane using Abell 2390.
For completeness, the `in-between' sample 
(between those considered to be group cluster and non-group cluster galaxies) is also plotted.
The \sig5~distribution can be described in general as having a decreasing
envelope with increasing radius, with high-density groups super-imposed on it.
The envelope distribution can be fitted reasonably well with a projected NFW \citep{1996ApJ...462..563N} profile.
It is interesting to note that Figure \ref{A2390.den} demonstrates
that the \sig5~measurements computed using photometric redshifts can produce a
reasonable projected galaxy density profile for the cluster, indicating that
these density measurements are in fact physically meaningful,
 at least when considered over a factor of several in density.

Figure \ref{gzfred} shows the redshift dependence of \fred~for the three subsamples of galaxies inside \rcl=3, using galaxies in all local galaxy density environments.
We see that all three samples show an increase of \fred~with decreasing 
redshift, but have systematically different levels of \fred, with
the cluster main-group being the reddest and the non-group cluster galaxies
the bluest.
Furthermore, the three subsamples also show different evolution in \fred.
The cluster main-group galaxies exhibit the largest $f_{red}$ and reach the 
final state of $f_{red}\sim 1$ earlier than the group cluster galaxies. 
The group cluster galaxies show the strongest evolution from $z\sim0.5$
to 0.2, reaching \fred~close to that of the cluster main-group galaxies
at the low-redshift bin.
The non-group cluster galaxies, located in less massive subhalos,
show a milder increase in \fred, still retaining a substantial fraction
of blue galaxies in the low-redshift bin. They
begin to partake in significant evolution at the lower redshift
of $z\sim0.35$.
In the context of the halo model, the cluster cores are the center
of the most massive dark matter halos, while group and non-group cluster
galaxies are in successively less massive subhalos. 
The epoch when $f_{red}$ reaches $\sim 1$ is therefore a reflection 
of the dependence of galaxy evolution on the halo mass: 
galaxies in more massive halos have their star formation truncated at an earlier time than do those in less massive halos. 
This result can be described as a `group down-sizing' effect,
akin to the down-sizing effect seen in the evolution of individual galaxies
\citep[e.g.,][]{1996AJ....112..839C,2008MNRAS.389..567C,2008MNRAS.386.1695S},
though the physical mechanisms underlying this effect may be very different.

Figure \ref{grfred} demonstrates the different evolutionary histories of
galaxies in subhalos of different sizes by plotting the dependence of \fred~on cluster centric radius for the three galaxy subsamples in different redshift bins.
We plot the cluster main-groups as a single point at a small radius.
Cluster galaxies in groups  show the largest increases in \fred~with decreasing
 redshift; and this increase is seen in both inner and outer
 cluster-centric radius bins. 
The \fred~for galaxies in groups in the cluster outskirts shows the largest change
with redshift, catching up to the value of the cluster main-group
galaxies in the core by $z\sim0.2$, by which time a flat \fred~distribution 
in cluster-centric radius is seen.
In comparison, the \fred~of non-group cluster galaxies shows a very strong
radial gradient at the low-redshift bin. 
The rate of evolution of \fred~for the non-group cluster galaxies
depends significantly on the cluster-centric radius, such that galaxies
in the outermost radial bin (at $\sim 2R_{200}$) show little change with 
redshift, while galaxies near $R_{200}$ evolve significantly.

Our results suggest that galaxies in moderately massive groups outside of
the cluster core are turning red in response to the environment of their
own dark matter subhalos, with the cluster environment having a strong
effect possibly only at the core.
On the other hand, the non-group cluster galaxies, situated in low
mass subhalos, appear to turn red at a low or non-existent rate at
the outskirts, and significant evolution is seen only when they are
near the virial radius of the cluster.
This result can be interpreted as galaxies in small subhalos being
much more affected by the massive cluster dark matter halo into which they
are falling.
However, another possible contribution to this increase of \fred~for
the non-group sample near the virial radius could come from member galaxies 
of more massive infalling groups (which have already turned red), 
which have been dynamically stripped from their parent subhalos.

To separate the effects of local galaxy density and group membership, we plot
in Figure \ref{gdfred}~\fred~for the three subsamples at different redshifts as a function of \sig5. 
The uncertainties are fairly large due to the small sample size, but,
nevertheless, the plots suggest that,
after controlling for the local galaxy density, 
the three subsamples still have different red fractions.
This result indicates that the different average local 
galaxy densities in these subsamples
do not entirely  explain the different \fred~measured.
Comparing galaxies in the cluster main-groups to those in groups in 
the high-redshift bin, the cluster galaxies in groups have substantially smaller \fred~at all densities, similar to non-group cluster galaxies.
By $z\sim0.2$,
the group cluster galaxies have evolved into having \fred~similar to the 
cluster main-group galaxies, while the non-group cluster galaxies show a 
considerably milder evolution.
Thus, by  $z\sim0.2$, cluster galaxies in groups are redder than 
their non-group counterparts in similar local galaxy densities, 
implying that group membership,
or more precisely, the mass of the subhalo in which a galaxy resides,
has a stronger effect on the evolution of its stellar population than does the local galaxy density.

\section{Discussion\label{discussion}}
The availability of photometric redshift information over a large field of view
allows us to examine the Butcher-Oemler effect in galaxy clusters in 
detail, including investigating the effects of local galaxy density and
the dependence on the global cluster scale.
Furthermore, by separating the cluster galaxies into different group status categories,
roughly indicative of the size of the dark matter subhalo in which they
reside, we are able to examine more closely the nature of the dependence
of the galaxy red fraction on their subhalo masses.

Figures \ref{rfred} to \ref{rplot} illustrate the apparent
dependence of galaxy colors on local galaxy density and cluster 
centric radius, averaged over the whole cluster galaxy sample.
Looking at cluster galaxies as a whole,
by $z\sim 0.20$, most cluster galaxies in high $\Sigma_5$ regions 
already have their star formation truncated, as their $f_{red}$ is 
close to 1 (see Figure 10).
The truncation process appears to occur earlier for galaxies in denser regions.
Assuming a galaxy would turn red 1$\sim$2 Gyr after star formation 
is quenched, this truncation process is mostly completed by the
redshift of $\sim0.45$ for galaxies in high-density regions, given 
that their $f_{red}$ reaches $\sim$1 by our low-redshift bin ($z\sim0.20$).

The cluster environment, as defined by the position of the galaxy in
the cluster, however, appears to have an even
 stronger effect on \fred~than the local galaxy density. 
For galaxies with similar $\Sigma_5$, those closer to the cluster 
center have a larger fraction of their star formation suppressed than those 
in the  outskirts (Figure 11, comparing within each row). 
When the galaxy sample is separated into subsamples in cluster-centric radius,
the dependence of \fred~on the local galaxy density 
is considerably weaker, and nearly non-existent
in the cores of clusters (Figure 12), suggesting that local galaxy
density is not the dominant environmental parameter.
We also find galaxies in low-density regions have a steeper
\fred-\rcl~gradient, compared to those in high-density regions
(Figure 13). This suggests that the effects of the global cluster
environment are stronger for galaxies in low-density regions.
However, an alternative interpretation might be that star formation 
in galaxies in high $\Sigma_5$ regions may have already been largely
stopped by some other processes not associated with the global 
cluster environment;  whereas in low $\Sigma_5$ regions, there is 
still a relatively large fraction of star-forming galaxies available
to have their star formation truncated by the cluster environment
as they approach the cluster virial region.

The weak dependence of \fred~on local galaxy density seen  when the data are
analyzed in different \rcl~bins (Figure \ref{dplot}) suggests that \sig5~may not be a   
fundamental parameter in determining the evolution of the galaxy population
in clusters.
Separating cluster galaxies into group and non-group galaxies, representing
galaxies in more and less massive subhalos, provides us with additional insights into cluster galaxy population evolution, and a more physical interpretation of the apparent dependence on local galaxy density.
 The dependence of $f_{red}$ on $\Sigma_5$ for both group and non-group  
cluster galaxies is also weak (Figure \ref{gdfred}), similar to that found for the (whole) cluster 
 galaxy sample after they are separated into bins of cluster-centric radii 
(Figure \ref{dplot}). 
The most significant result from applying the group and non-group
selection of galaxy subsamples is presented in Figure \ref{grfred}.
It presents clear evidence that, outside the cluster core, galaxies
situated in the more massive subhalos (the group cluster galaxies) have
a much stronger Butcher-Oemler effect, regardless of their position relative to the cluster center.
In the context of a dark matter halo model, this suggests that much of the dependence of the galaxy population on local galaxy density can be attributed to whether an infalling cluster galaxy is in a massive subhalo.

Our analysis shows that cluster galaxies in groups, regardless of their local
 galaxy density and cluster-centric radius, exhibit a much stronger
 Butcher-Oemler effect than the non-group cluster galaxies (Figure \ref{grfred}
 and \ref{gdfred}); the difference between
 these subsamples is especially large beyond the virial radius.
In fact, the cluster galaxies considered to be in moderately massive 
subhalos appear to be the most significant contributors to the 
observed Butcher-Oemler effect at the cluster outskirts ($\sim2-3 R_{200}$)
over the redshift range of our sample.
This dependence on halo mass can also be extended to include the core 
of a cluster, which is the center of the most massive halo.
The cores of the clusters have high red fractions close to 1
at $z\sim0.4$;
 they  complete the quenching of star formation
the earliest, ahead of the less massive infalling subhalos.

The evolution of \fred~of the large sample of cluster galaxies 
with pFoF group membership
falling in between the group cluster and non-group cluster galaxies
is also consistent with this picture.
This sample contains the bulk of the cluster galaxies --
 about 49\% of the cluster galaxies outside the cluster core.
The \fred~of this sample at the outer cluster-centric radial bins
show a Butcher-Oemler effect that is between those of the
group cluster and non-group cluster galaxies.
For example, at the outermost $r_{CL}$ bin,  \fred~rises from $0.19\pm0.02$ at
$z\sim0.4$ to $0.42\pm0.06$ at $z\sim0.2$.
It is more difficult to quantify the halo mass of objects in this sample 
because of the high false group detection rate (see \S8.4).
This sample is expected to contain a broad mix of subhalo masses,
with very roughly half of the galaxies being in subhalos with masses of
$\sim 1-3\times 10^{13}$\msun, and the other half, being contaminated by 
projections, residing in subhalos of smaller masses, similar to those
in the non-group cluster galaxy sample.
The more massive subhalos in this sample, along with the group cluster
galaxy sample, could represent a total of  30 to 45\%~of all the
the infalling cluster galaxies. They are likely the main contributors
to Butcher-Oemler effect seen in the outer radial bins.

This rapid evolution of group populations at $z\sim0.3$, combined with the
red cores of clusters seen at higher redshifts and the bluer colors
of galaxies in smaller groups, implies that
galaxies in the most massive halos (i.e., in the cluster cores) have star formation quenched at an earlier time than those in moderately massive infalling groups.
These, in turn, become red at an earlier time than the non-group galaxies 
residing in yet less massive subhalos (Figure 12).
This effect suggests that mechanisms responsible for the truncation
of star formation in galaxies in more massive halos operate more 
efficiently, or become more effective in earlier epochs.

Related to the more rapid evolution of \fred~in group galaxies
is the `pre-processing' effect on galaxies in infalling groups, suggested by number of previous studies \citep[e.g.,][]{2004MNRAS.355..713B,2005MNRAS.362..268T,2008MNRAS.tmp.1037C,2008MNRAS.388.1537M}.
We clearly observe this in Figure \ref{grfred}, as the \fred~of galaxies 
in groups increases with decreasing redshift at all cluster-centric radii.
The pre-processing effect is essentially a manifestation that the more massive subhalos in a cluster have a dominant effect on their galaxy populations,
producing a strong Butcher-Oemler effect regardless of their distance from the cluster center. 
By $z\sim0.35$, this `processing' of galaxies 
in moderately massive subhalos results in galaxies which are
redder than the non-group galaxies in similar local galaxy density 
regions located at similar cluster-centric radii.
The \fred~of these galaxies, even at large \rcl, becomes similar 
to that of the evolved galaxies in the cluster core by $z\sim0.2$.
While our sample demonstrates this effect in groups with subhalo masses
larger than $\sim3-4\times 10^{13}$ \msun, 
it also likely occurs in subhalos 2 or 3 times less massive.
This increase in the dominance of red galaxies in groups appears to be a significant, if not the primary, contributor of the Butcher-Oemler effect over the redshift range of 0.5 to 0.2.
The increase of \fred~for the galaxies in both the non-group cluster galaxies and the cluster core galaxies is considerably smaller over this redshift range,
in that the galaxies in the cluster core have already completed their
evolution, and small subhalos (hosting the non-group cluster
galaxies) outside the core still contain a substantial number of blue
galaxies even at $z\sim0.2$.

\section{Conclusions \label{conclusions}}
\begin{figure}
	\includegraphics[width=15cm]{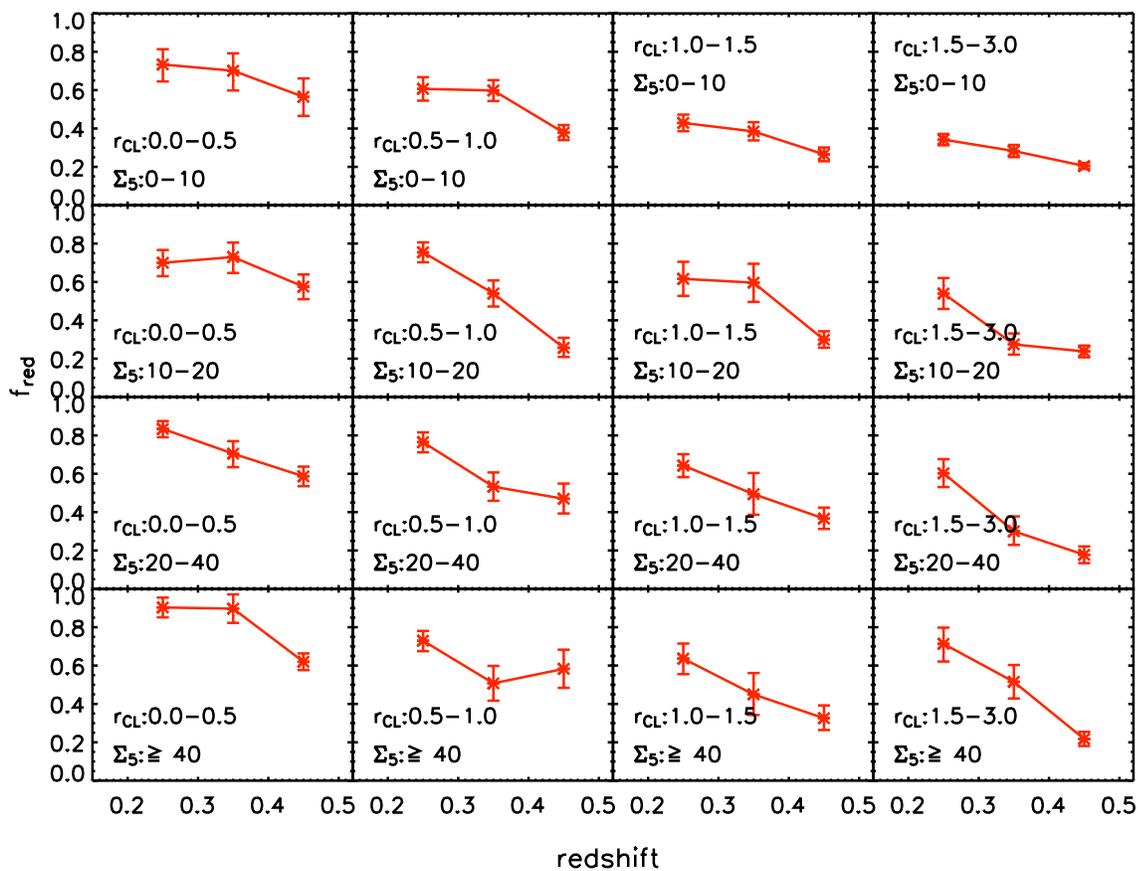}
	\caption{The \it `$f_{red}$-z' \rm trends in different environments, where $r_{CL}$ increases from the left to the right and $\Sigma_5$ increases from the top to the bottom. The general trend of $f_{red}$ decreasing with increasing redshift is seen in all panels. \label{zplot}}
	\end{figure}

\begin{figure}
	\includegraphics[width=15cm]{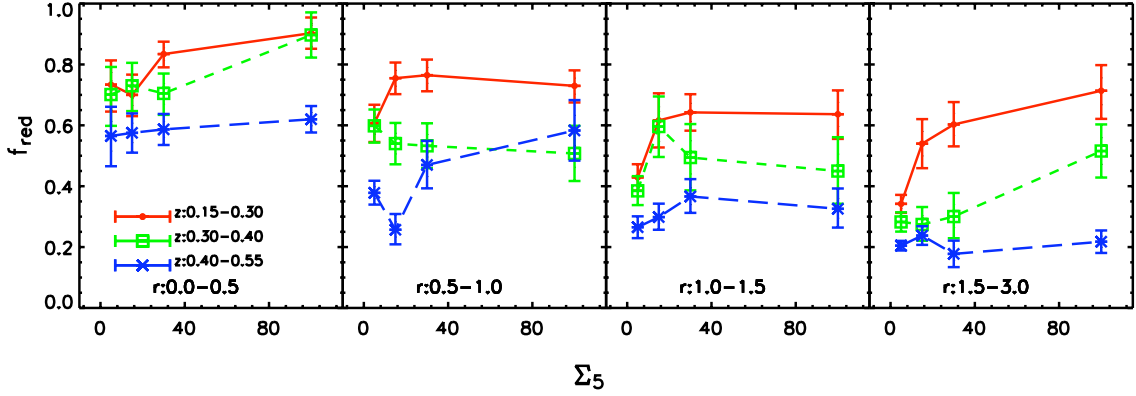}
	\caption{The \it `$f_{red}$-$\Sigma_5$' \rm trends in different $r_{CL}$ and redshift bins. 
The trend of $f_{red}$ increasing with $\Sigma_5$ is relatively weak 
compared to  those in Figure \ref{dfred}.
This trend is most apparent at low redshift and at large $r_{CL}$.  \label{dplot}}
	\end{figure}

\begin{figure}
	\includegraphics[width=15cm]{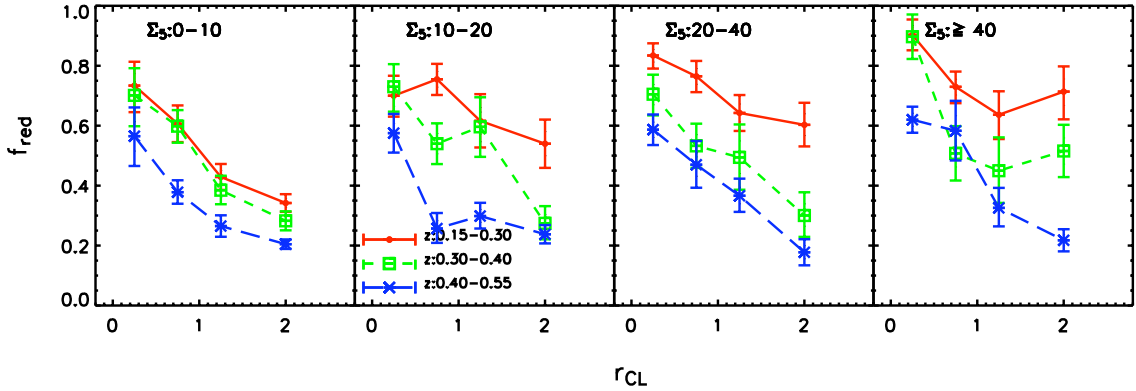}
	\caption{The \it `$f_{red}$-$r_{CL}$' \rm trends in different $\Sigma_5$ and redshift bins. 
The trend with $r_{CL}$ is strongest in the $\Sigma_5$=0--10 bin
and for the high-redshift subsample in the other $\Sigma_5$~bins. \label{rplot}}
	\end{figure}

\begin{figure}
	\includegraphics{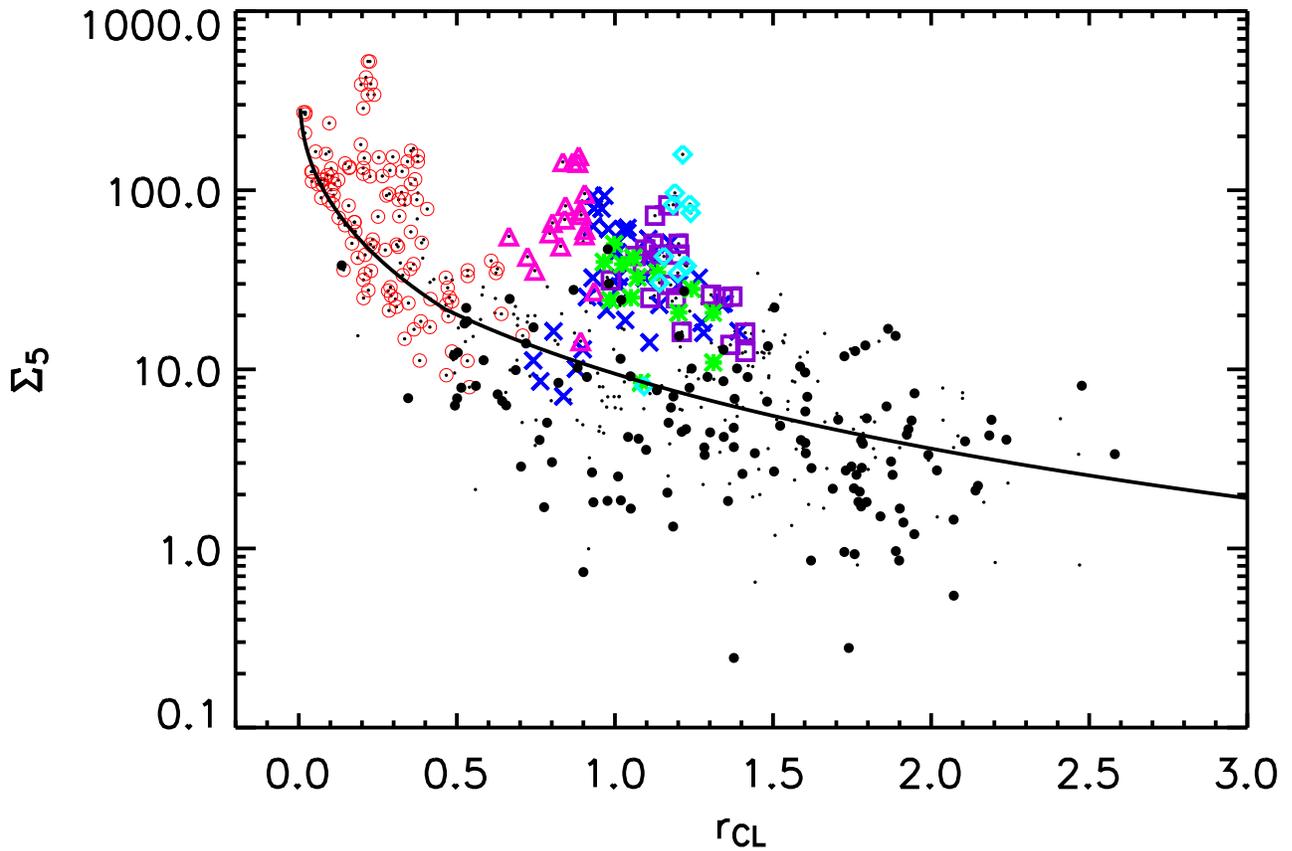}
	\caption{Local galaxy density $\Sigma_5$ as a function of cluster-centric radius $r_{CL}$ in units of $R_{200}$ using Abell 2390  cluster galaxies. 
The crosses, squares, triangles, stars, and diamonds represent group galaxies in different cluster groups (see Figure \ref{A2390.map}). 
The thick solid circles are the `non-group' cluster galaxies and those marked with large open circles are the cluster galaxies in the cluster main-group.
The small dots are those in between our `group' and 'non-group' categories. 
In general, the group galaxies populate regions 
of $\Sigma_5$ relatively higher than
those of the non-group galaxies.
The solid curve is the projected NFW profile fitted using the cluster main-group and non-group galaxies.
	\label{A2390.den}}
	\end{figure}

\begin{figure}
	\includegraphics{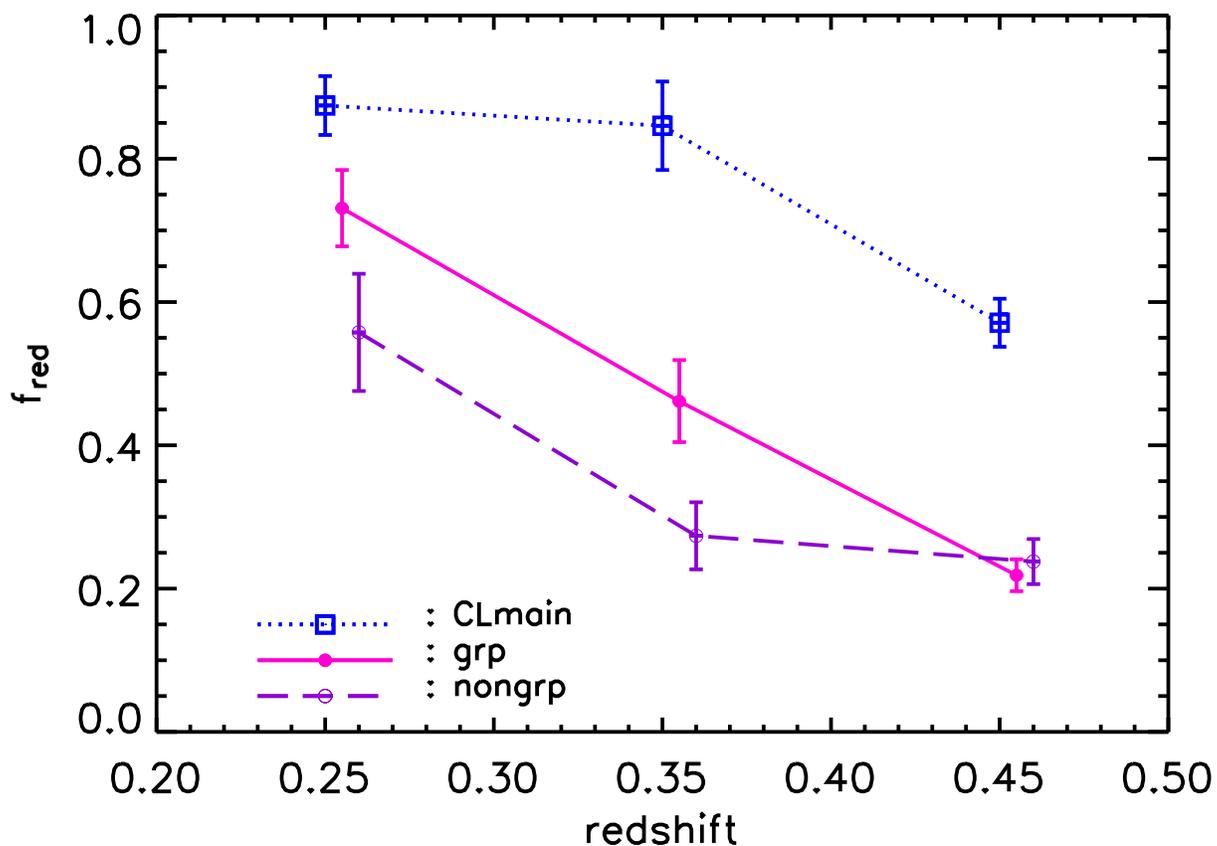}
	\caption{The \it `$f_{red}$-z' \rm trends for the cluster main-group, 
group cluster, and non-group cluster galaxies as indicated on the plot,
using galaxies in all $\Sigma_5$ bins. All group and non-group galaxies are located at $r_{CL} < 3$.
The cluster main-group subsample has the largest $f_{red}$, while the non-group galaxies exhibit a weak \it `$f_{red}$-z' \rm trend. \label{gzfred}}
	\end{figure}

\begin{figure}
	\includegraphics{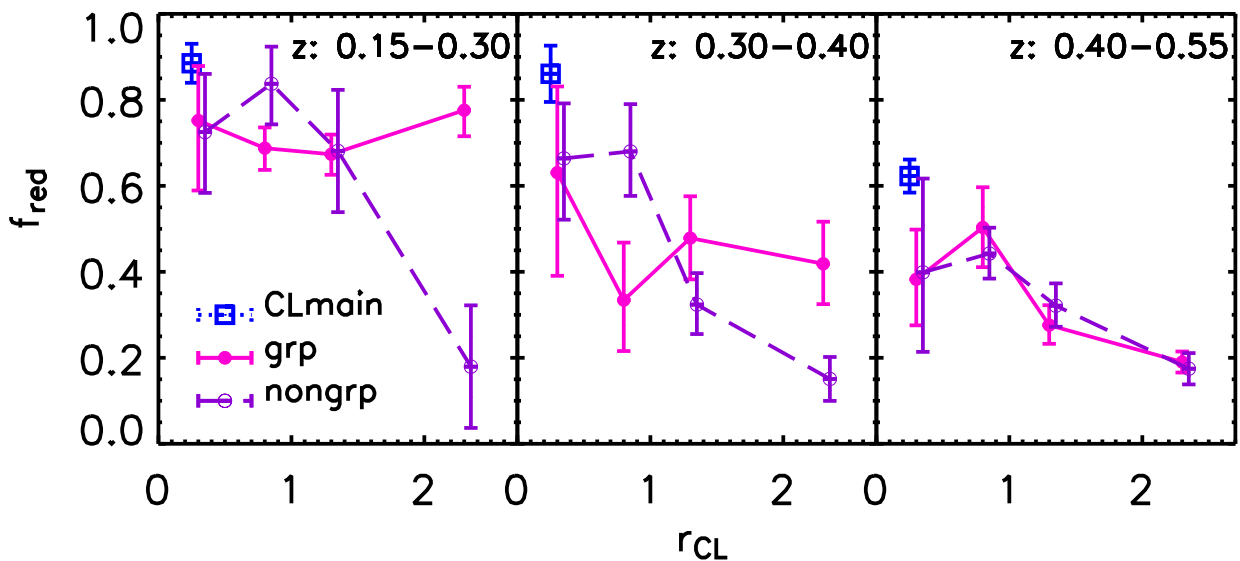}
	\caption{The \it `$f_{red}$-$r_{CL}$' \rm trends in different redshift bins for the three subsamples indicated on the plot.
Group cluster galaxies outside $\sim 1 R_{200}$ evolve much more rapidly than  non-group cluster galaxies over this redshift range.
\label{grfred}}
	\end{figure}

\begin{figure}
	\includegraphics{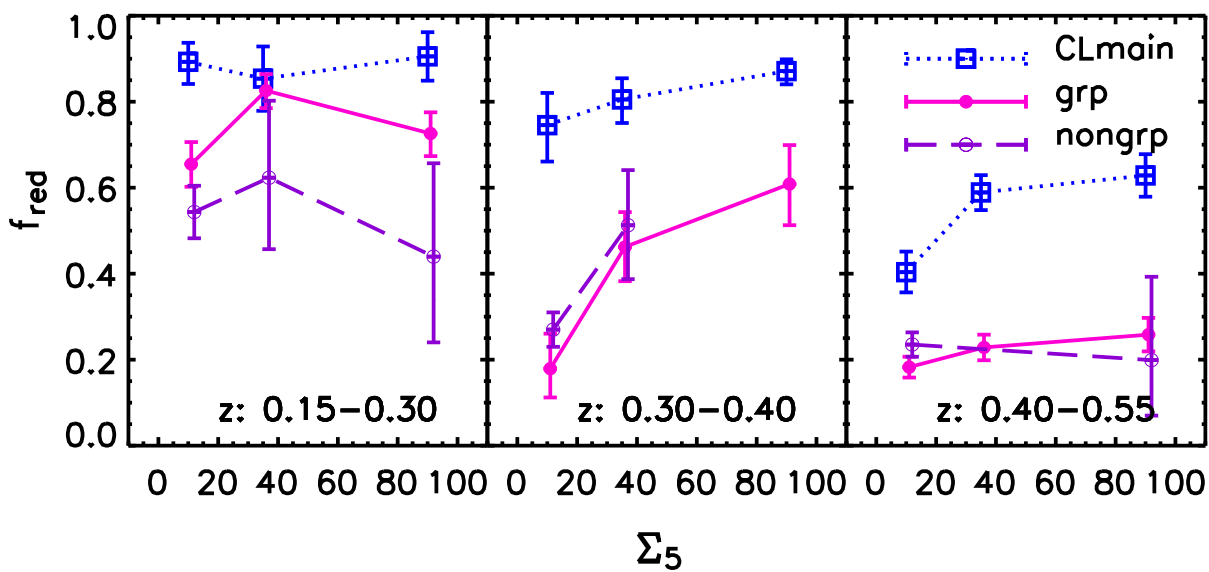}
\caption{The \it `$f_{red}$-$\Sigma_5$' \rm trends in different redshift bins for the three subsamples indicated on the plot.
\label{gdfred}}
	\end{figure}

 	We have investigated the environmental effects on the properties of the CNOC1 cluster galaxies at $0.17<z<0.55$ using both 
individual galaxies and those in galaxy groups,
utilizing photometric redshift data covering a relatively large field out to 2 -- 3 $R_{200}$.
We use local projected galaxy density, $\Sigma_5$, and cluster-centric 
radius, $r_{CL}$, in units of $R_{200}$ as proxies for the  local and 
global environment, and adopt the red galaxy fraction, $f_{red}$, 
as the measurement of galaxy population properties. 
We summarize our results as follows:

\noindent
1.) The Butcher-Oemler effect is observed in all $r_{CL}$ and $\Sigma_5$ bins between redshifts of 0.5 and 0.2 (Figure \ref{zplot}). 
However, the change of $f_{red}$ in the cluster core (less than 0.5 $R_{200}$) is minimal at $z<0.4$, suggesting that the truncation of star formation in the center of the most massive halos occurred at earlier epochs.
 
\noindent
2.) We find that stellar-mass selected samples also show the Butcher-Oemler
effect. 
Comparing the samples selected using different stellar-mass limits 
(at 1 $R_{200}$), we find the steepness of the Butcher-Oemler effect to be
 dependent on
the stellar-mass limit. A sample selected using a higher stellar-mass
limit has a milder change in \fred~compared to one with a lower stellar-mass
limit.

\noindent
3.) There are apparent correlations between $\Sigma_5$ 
and $f_{red}$ in all our three redshift bins. 
However, the dependence of \fred~on local galaxy density 
becomes much weaker when the galaxy sample is separated by 
cluster-centric radius or group/non-group status, with the possible 
exception at the cluster outskirts. This suggests that local
galaxy density may not be the primary environment parameter
affecting the average galaxy population.

\noindent
4.)
We separate the cluster galaxies into three subsamples: cluster main-group, 
group cluster, and non-group cluster galaxies, roughly representing galaxies
situated in the core of the massive dark matter halo of the cluster,
in moderately massive subhalos, and in low-mass subhalos, respectively.
These three samples have different \fred~values and rates of change
of \fred~with redshift. 
Galaxies in the central massive cluster halo reach $f_{red}\sim1$ 
at a larger redshift than those in galaxy groups, which 
in turn reach $f_{red}\sim1$ at an earlier time than the non-group galaxies.    
This suggests that the dominant determinant of the epoch of the quenching of 
star formation is the mass of the dark matter subhalos in which the galaxy resides. 
The apparent correlation between $\Sigma_5$ and $f_{red}$ is at least 
in part due to the correlation of high local galaxy density with the presence 
of a massive halo.    
The epoch when \fred~in a galaxy group reaches the evolved stage of  $f_{red}\sim1$ depends on the halo mass. 

\noindent
5.) We find that galaxy groups falling into a cluster `process' galaxies 
therein and turn them red at an earlier time; what is termed as `pre-processing'. 
This is observed even after controlling for $\Sigma_5$.
Such `pre-processing' is seen for groups at the outer $r_{CL}$ and 
the cluster outskirts. 
By $z\sim0.20$, galaxies in groups have a high red fraction,
similar to that in the cluster core, regardless of their cluster-centric 
radius.                                    
This more rapid evolution of galaxies in infalling moderately rich
groups appears to be a significant contributor to the Butcher-Oemler effect.

Some of our analyses and interpretations are tentative and suggestive, based on a small-size sample of 16 clusters and 89 cluster groups. 
The completion of the \it Red-sequence Cluster Survey \rm 1 and 2
\citep[RCS;][]{2005ApJS..158..161H,2007ASPC..379..103Y}
promises to provide similar data for thousands of clusters,
providing greatly improved statistics to investigate the roles of
dark matter halo environment on galaxy evolution.

\acknowledgements
This paper is
based on observations obtained at the Canada-France-Hawaii
Telescope (CFHT), which is operated by the National Research Council
(NRC) of Canada, the Institut National des Sciences de l'Univers of the
Centre National de la Recherche Scientifique (CNRS) of France, and
the University of Hawaii.

We sincerely thank David Gilbank for helpful discussions. 
We also thank the anonymous referee for the valuable points of using stellar-mass sample.
I.H.L. acknowledges financial supports from the University of Toronto Fellowship, the Helen Hogg Fellowship, and grants to H.K.C.Y. from the National Science Engineering Research Council (NSERC). 
The research of H.K.C.Y. is supported by grants from the Canada 
Research Chair Program, NSERC and the University of Toronto. E. E. acknowledges NSF grant AST0206154 for support.


\begin{thebibliography}{77}
\expandafter\ifx\csname natexlab\endcsname\relax\def\natexlab#1{#1}\fi

\bibitem[{{Abell} {et~al.}(1989){Abell}, {Corwin}, \&
  {Olowin}}]{1989ApJS...70....1A}
{Abell}, G.~O., {Corwin}, Jr., H.~G., \& {Olowin}, R.~P. 1989, \apjs, 70, 1

\bibitem[{{Abraham} {et~al.}(1996)}]{1996ApJ...471..694A}
{Abraham}, R.~G., et al. 1996, \apj, 471, 694

\bibitem[{{Adami} {et~al.}(2005){Adami}, {Biviano}, {Durret}, \&
  {Mazure}}]{2005A&A...443...17A}
{Adami}, C., {Biviano}, A., {Durret}, F., \& {Mazure}, A. 2005, \aap, 443, 17

\bibitem[{{Andreon} \& {Ettori}(1999)}]{1999ApJ...516..647A}
{Andreon}, S. \& {Ettori}, S. 1999, \apj, 516, 647

\bibitem[{{Andreon} {et~al.}(2004){Andreon}, {Lobo}, \&
  {Iovino}}]{2004MNRAS.349..889A}
{Andreon}, S., {Lobo}, C., \& {Iovino}, A. 2004, \mnras, 349, 889

\bibitem[{{Andreon} {et~al.}(2006){Andreon}, {Quintana}, {Tajer}, {Galaz}, \&
  {Surdej}}]{2006MNRAS.365..915A}
{Andreon}, S., {Quintana}, H., {Tajer}, M., {Galaz}, G., \& {Surdej}, J. 2006,
  \mnras, 365, 915

\bibitem[{{Barkhouse} {et~al.}(2007){Barkhouse}, {Yee}, \&
  {L{\'o}pez-Cruz}}]{2007ApJ...671.1471B}
{Barkhouse}, W.~A., {Yee}, H.~K.~C., \& {L{\'o}pez-Cruz}, O. 2007, \apj, 671,
  1471

\bibitem[{{Bekki}(1999)}]{1999ApJ...510L..15B}
{Bekki}, K. 1999, \apjl, 510, L15

\bibitem[{{Bell} {et~al.}(2003){Bell}, {McIntosh}, {Katz}, \&
  {Weinberg}}]{2003ApJS..149..289B}
{Bell}, E.~F., {McIntosh}, D.~H., {Katz}, N., \& {Weinberg}, M.~D. 2003, \apjs,
  149, 289

\bibitem[{{Berrier} {et~al.}(2009){Berrier}, {Stewart}, {Bullock}, {Purcell},
  {Barton}, \& {Wechsler}}]{2009ApJ...690.1292B}
{Berrier}, J.~C., {Stewart}, K.~R., {Bullock}, J.~S., {Purcell}, C.~W.,
  {Barton}, E.~J., \& {Wechsler}, R.~H. 2009, \apj, 690, 1292

\bibitem[{{Blake} {et~al.}(2004)}]{2004MNRAS.355..713B}
 {Blake}, C., et al. 2004, \mnras, 355, 713

\bibitem[{{Blanton} {et~al.}(2005){Blanton}, {Eisenstein}, {Hogg}, {Schlegel},
  \& {Brinkmann}}]{2005ApJ...629..143B}
{Blanton}, M.~R., {Eisenstein}, D., {Hogg}, D.~W., {Schlegel}, D.~J., \&
  {Brinkmann}, J. 2005, \apj, 629, 143

\bibitem[{{Borgani} {et~al.}(1999){Borgani}, {Girardi}, {Carlberg}, {Yee}, \&
  {Ellingson}}]{1999ApJ...527..561B}
{Borgani}, S., {Girardi}, M., {Carlberg}, R.~G., {Yee}, H.~K.~C., \&
  {Ellingson}, E. 1999, \apj, 527, 561

\bibitem[{{Butcher} \& {Oemler}(1984)}]{1984ApJ...285..426B}
{Butcher}, H. \& {Oemler}, Jr., A. 1984, \apj, 285, 426

\bibitem[{{Carlberg} {et~al.}(1996){Carlberg}, {Yee}, {Ellingson}, {Abraham},
  {Gravel}, {Morris}, \& {Pritchet}}]{1996ApJ...462...32C}
{Carlberg}, R.~G., {Yee}, H.~K.~C., {Ellingson}, E., {Abraham}, R., {Gravel},
  P., {Morris}, S., \& {Pritchet}, C.~J. 1996, \apj, 462, 32

\bibitem[{{Carlberg} {et~al.}(1997)}]{1997ApJ...485L..13C}
{Carlberg}, R.~G., et al. 1997, \apjl, 485, L13+

\bibitem[{{Cattaneo} {et~al.}(2008){Cattaneo}, {Dekel}, {Faber}, \&
  {Guiderdoni}}]{2008MNRAS.389..567C}
{Cattaneo}, A., {Dekel}, A., {Faber}, S.~M., \& {Guiderdoni}, B. 2008, \mnras,
  389, 567

\bibitem[{{Cooper} {et~al.}(2007)}]{2007MNRAS.tmp..224C}
{Cooper}, M.~C., et al. 2007, \mnras, 224

\bibitem[{{Cooper} {et~al.}(2008){Cooper}, {Tremonti}, {Newman}, \&
  {Zabludoff}}]{2008MNRAS.tmp.1037C}
{Cooper}, M.~C., {Tremonti}, C.~A., {Newman}, J.~A., \& {Zabludoff}, A.~I.
  2008, \mnras, 1037

\bibitem[{{Cowie} {et~al.}(1996){Cowie}, {Songaila}, {Hu}, \&
  {Cohen}}]{1996AJ....112..839C}
{Cowie}, L.~L., {Songaila}, A., {Hu}, E.~M., \& {Cohen}, J.~G. 1996, \aj, 112,
  839

\bibitem[{{Coziol} \& {Plauchu-Frayn}(2007)}]{2007AJ....133.2630C}
{Coziol}, R. \& {Plauchu-Frayn}, I. 2007, \aj, 133, 2630

\bibitem[{{D'Agostini}(2004)}]{2004physics..12069D}
{D'Agostini}, G. 2004, ArXiv Physics e-prints

\bibitem[{{Dahl{\'e}n} {et~al.}(2004){Dahl{\'e}n}, {Fransson}, {{\"O}stlin}, \&
  {N{\"a}slund}}]{2004MNRAS.350..253D}
{Dahl{\'e}n}, T., {Fransson}, C., {{\"O}stlin}, G., \& {N{\"a}slund}, M. 2004,
  \mnras, 350, 253

\bibitem[{{David} {et~al.}(1999){David}, {Forman}, \&
  {Jones}}]{1999ApJ...519..533D}
{David}, L.~P., {Forman}, W., \& {Jones}, C. 1999, \apj, 519, 533

\bibitem[{{De Lucia} {et~al.}(2004)}]{2004ApJ...610L..77D}
{De Lucia}, G., et al. 2004, \apjl, 610, L77

\bibitem[{{De Lucia} {et~al.}(2007)}]{2007MNRAS.374..809D}
{De Lucia}, G., et al. 2007, \mnras, 374, 809

\bibitem[{{De Lucia} {et~al.}(2006){De Lucia}, {Springel}, {White}, {Croton},
  \& {Kauffmann}}]{2006MNRAS.366..499D}
{De Lucia}, G., {Springel}, V., {White}, S.~D.~M., {Croton}, D., \&
  {Kauffmann}, G. 2006, \mnras, 366, 499

\bibitem[{{De Propris} {et~al.}(2004)}]{2004MNRAS.351..125D}
{De Propris}, R., 2004, \mnras, 351, 125

\bibitem[{{De Propris} {et~al.}(2003){De Propris}, {Stanford}, {Eisenhardt}, \&
  {Dickinson}}]{2003ApJ...598...20D}
{De Propris}, R., {Stanford}, S.~A., {Eisenhardt}, P.~R., \& {Dickinson}, M.
  2003, \apj, 598, 20

\bibitem[{{Dom{\'{\i}}nguez} {et~al.}(2002){Dom{\'{\i}}nguez}, {Zandivarez},
  {Mart{\'{\i}}nez}, {Merch{\'a}n}, {Muriel}, \&
  {Lambas}}]{2002MNRAS.335..825D}
{Dom{\'{\i}}nguez}, M.~J., {Zandivarez}, A.~A., {Mart{\'{\i}}nez}, H.~J.,
  {Merch{\'a}n}, M.~E., {Muriel}, H., \& {Lambas}, D.~G. 2002, \mnras, 335, 825

\bibitem[{{Dressler}(1980)}]{1980ApJS...42..565D}
{Dressler}, A. 1980, \apjs, 42, 565

\bibitem[{{Ellingson} {et~al.}(2001){Ellingson}, {Lin}, {Yee}, \&
  {Carlberg}}]{2001ApJ...547..609E}
{Ellingson}, E., {Lin}, H., {Yee}, H.~K.~C., \& {Carlberg}, R.~G. 2001, \apj,
  547, 609

\bibitem[{{Fujita}(2004)}]{2004PASJ...56...29F}
{Fujita}, Y. 2004, \pasj, 56, 29

\bibitem[{{Gerke} {et~al.}(2007)}]{2007MNRAS.tmp..222G}
{Gerke}, B.~F., et al. 2007, \mnras, 222

\bibitem[{{Giavalisco} {et~al.}(2004)}]{2004ApJ...600L..93G}
{Giavalisco}, M., et al. 2004, \apjl, 600, L93

\bibitem[{{Gilbank} {et~al.}(2008){Gilbank}, {Yee}, {Ellingson},
 {Gladders}, {Loh}, {Barrientos}, \& {Barkhouse}}]{2008ApJ...673..742G}
 {Gilbank}, D.~G., {Yee}, H.K.C., {Ellingson}, E., {Gladders}, M.~D., {Loh},
 Y.-S., {Barrientos}, L.~F., \& {Barkhouse}, W.~A. 2008, \apj, 673, 74

\bibitem[{{Gladders} \& {Yee}(2005)}]{2005ApJS..157....1G}
{Gladders}, M.~D. \& {Yee}, H.~K.~C. 2005, \apjs, 157, 1

\bibitem[{{G{\'o}mez} {et~al.}(2003)}]{2003ApJ...584..210G}
{G{\'o}mez}, P.~L., et al. 2003, \apj, 584, 210

\bibitem[{{Goto} {et~al.}(2003){Goto}, {Yamauchi}, {Fujita}, {Okamura},
  {Sekiguchi}, {Smail}, {Bernardi}, \& {Gomez}}]{2003MNRAS.346..601G}
{Goto}, T., {Yamauchi}, C., {Fujita}, Y., {Okamura}, S., {Sekiguchi}, M.,
  {Smail}, I., {Bernardi}, M., \& {Gomez}, P.~L. 2003, \mnras, 346, 601

\bibitem[{{Helsdon} \& {Ponman}(2003{\natexlab{a}})}]{2003MNRAS.339L..29H}
{Helsdon}, S.~F. \& {Ponman}, T.~J. 2003{\natexlab{a}}, \mnras, 339, L29

\bibitem[{{Helsdon} \& {Ponman}(2003{\natexlab{b}})}]{2003MNRAS.340..485H}
---. 2003{\natexlab{b}}, \mnras, 340, 485

\bibitem[{{Hogg} {et~al.}(2003)}]{2003ApJ...585L...5H}
{Hogg}, D.~W., et al. 2003, \apjl, 585,
  L5

\bibitem[{{Holden} {et~al.}(2007)}]{2007ApJ...670..190H}
{Holden}, B.~P., et al. 2007, \apj, 670, 190

\bibitem[{{Hsieh} {et~al.}(2005){Hsieh}, {Yee}, {Lin}, \&
  {Gladders}}]{2005ApJS..158..161H}
{Hsieh}, B.~C., {Yee}, H.~K.~C., {Lin}, H., \& {Gladders}, M.~D. 2005, \apjs,
  158, 161

\bibitem[{{Jeltema} {et~al.}(2007){Jeltema}, {Mulchaey}, {Lubin}, \&
  {Fassnacht}}]{2007ApJ...658..865J}
{Jeltema}, T.~E., {Mulchaey}, J.~S., {Lubin}, L.~M., \& {Fassnacht}, C.~D.
  2007, \apj, 658, 865

\bibitem[{{Kodama} \& {Arimoto}(1997)}]{1997A&A...320...41K}
{Kodama}, T. \& {Arimoto}, N. 1997, \aap, 320, 41

\bibitem[{{Kodama} {et~al.}(2001){Kodama}, {Smail}, {Nakata}, {Okamura}, \&
  {Bower}}]{2001ApJ...562L...9K}
{Kodama}, T., {Smail}, I., {Nakata}, F., {Okamura}, S., \& {Bower}, R.~G. 2001,
  \apjl, 562, L9

\bibitem[{{Kodama} {et~al.}(2004)}]{2004MNRAS.350.1005K}
{Kodama}, T., et al. 2004, \mnras, 350, 1005

\bibitem[{{Krick} {et~al.}(2006){Krick}, {Bernstein}, \&
  {Pimbblet}}]{2006AJ....131..168K}
{Krick}, J.~E., {Bernstein}, R.~A., \& {Pimbblet}, K.~A. 2006, \aj, 131, 168

\bibitem[{{Landolt}(1992)}]{1992AJ....104..340L}
{Landolt}, A.~U. 1992, \aj, 104, 340

\bibitem[{{Li} \& {Yee}(2008)}]{2008AJ....135..809L}
{Li}, I.~H. \& {Yee}, H.~K.~C. 2008, \aj, 135, 809

\bibitem[{{Lilly} {et~al.}(2003){Lilly}, {Carollo}, \&
  {Stockton}}]{2003ApJ...597..730L}
{Lilly}, S.~J., {Carollo}, C.~M., \& {Stockton}, A.~N. 2003, \apj, 597, 730

\bibitem[{{Lin} {et~al.}(1999){Lin}, {Yee}, {Carlberg}, {Morris}, {Sawicki},
  {Patton}, {Wirth}, \& {Shepherd}}]{1999ApJ...518..533L}
{Lin}, H., {Yee}, H.~K.~C., {Carlberg}, R.~G., {Morris}, S.~L., {Sawicki}, M.,
  {Patton}, D.~R., {Wirth}, G., \& {Shepherd}, C.~W. 1999, \apj, 518, 533

\bibitem[{{Loh} {et~al.}(2008){Loh}, {Ellingson}, {Yee}, {Gilbank}, {Gladders},
  \& {Barrientos}}]{2008ApJ...680..214L}
{Loh}, Y.-S., {Ellingson}, E., {Yee}, H.~K.~C., {Gilbank}, D.~G., {Gladders},
  M.~D., \& {Barrientos}, L.~F. 2008, \apj, 680, 214

\bibitem[{{Mamon}(2000)}]{2000ASPC..197..377M}
{Mamon}, G.~A. 2000, in Astronomical Society of the Pacific Conference Series,
  Vol. 197, Dynamics of Galaxies: from the Early Universe to the Present, ed.
  F.~{Combes}, G.~A. {Mamon}, \& V.~{Charmandaris}, 377--+

\bibitem[{{Margoniner} {et~al.}(2001){Margoniner}, {de Carvalho}, {Gal}, \&
  {Djorgovski}}]{2001ApJ...548L.143M}
{Margoniner}, V.~E., {de Carvalho}, R.~R., {Gal}, R.~R., \& {Djorgovski}, S.~G.
  2001, \apjl, 548, L143

\bibitem[{{McIntosh} {et~al.}(2008){McIntosh}, {Guo}, {Hertzberg}, {Katz},
  {Mo}, {van den Bosch}, \& {Yang}}]{2008MNRAS.388.1537M}
{McIntosh}, D.~H., {Guo}, Y., {Hertzberg}, J., {Katz}, N., {Mo}, H.~J., {van
  den Bosch}, F.~C., \& {Yang}, X. 2008, \mnras, 388, 1537

\bibitem[{{Metevier} {et~al.}(2000){Metevier}, {Romer}, \&
  {Ulmer}}]{2000AJ....119.1090M}
{Metevier}, A.~J., {Romer}, A.~K., \& {Ulmer}, M.~P. 2000, \aj, 119, 1090

\bibitem[{{Mulchaey} \& {Zabludoff}(1998)}]{1998ApJ...496...73M}
{Mulchaey}, J.~S. \& {Zabludoff}, A.~I. 1998, \apj, 496, 73

\bibitem[{{Navarro} {et~al.}(1996){Navarro}, {Frenk}, \&
  {White}}]{1996ApJ...462..563N}
{Navarro}, J.~F., {Frenk}, C.~S., \& {White}, S.~D.~M. 1996, \apj, 462, 563

\bibitem[{{Nuijten} {et~al.}(2005){Nuijten}, {Simard}, {Gwyn}, \&
  {R{\"o}ttgering}}]{2005ApJ...626L..77N}
{Nuijten}, M.~J.~H.~M., {Simard}, L., {Gwyn}, S., \& {R{\"o}ttgering}, H.~J.~A.
  2005, \apjl, 626, L77

\bibitem[{{Pierre} {et~al.}(1996){Pierre}, {Le Borgne}, {Soucail}, \&
  {Kneib}}]{1996A&A...311..413P}
{Pierre}, M., {Le Borgne}, J.~F., {Soucail}, G., \& {Kneib}, J.~P. 1996, \aap,
  311, 413

\bibitem[{{Poggianti}(1997)}]{1997A&AS..122..399P}
{Poggianti}, B.~M. 1997, \aaps, 122, 399

\bibitem[{{Poggianti} {et~al.}(2006)}]{2006ApJ...642..188P}
{Poggianti}, B.~M., et al. 2006, \apj, 642, 188

\bibitem[{{Seymour} {et~al.}(2008)}]{2008MNRAS.386.1695S}
{Seymour}, N., et al. 2008, \mnras, 386, 1695

\bibitem[{{Solanes} {et~al.}(1999){Solanes}, {Salvador-Sol{\'e}}, \&
  {Gonz{\'a}lez-Casado}}]{1999A&A...343..733S}
{Solanes}, J.~M., {Salvador-Sol{\'e}}, E., \& {Gonz{\'a}lez-Casado}, G. 1999,
  \aap, 343, 733

\bibitem[{{Springel et al.}(2005)}]{2005Natur.435..629S}
{Springel et al.} 2005, \nat, 435, 629

\bibitem[{{Tanaka} {et~al.}(2005){Tanaka}, {Kodama}, {Arimoto}, {Okamura},
  {Umetsu}, {Shimasaku}, {Tanaka}, \& {Yamada}}]{2005MNRAS.362..268T}
{Tanaka}, M., {Kodama}, T., {Arimoto}, N., {Okamura}, S., {Umetsu}, K.,
  {Shimasaku}, K., {Tanaka}, I., \& {Yamada}, T. 2005, \mnras, 362, 268

\bibitem[{{Tran} {et~al.}(2005){Tran}, {van Dokkum}, {Illingworth}, {Kelson},
  {Gonzalez}, \& {Franx}}]{2005ApJ...619..134T}
{Tran}, K.-V.~H., {van Dokkum}, P., {Illingworth}, G.~D., {Kelson}, D.,
  {Gonzalez}, A., \& {Franx}, M. 2005, \apj, 619, 134

\bibitem[{{Ulmer} \& {Cruddace}(1982)}]{1982ApJ...258..434U}
{Ulmer}, M.~P. \& {Cruddace}, R.~G. 1982, \apj, 258, 434

\bibitem[{{van den Bosch} {et~al.}(2008){van den Bosch}, {Aquino}, {Yang},
  {Mo}, {Pasquali}, {McIntosh}, {Weinmann}, \& {Kang}}]{2008MNRAS.387...79V}
{van den Bosch}, F.~C., {Aquino}, D., {Yang}, X., {Mo}, H.~J., {Pasquali}, A.,
  {McIntosh}, D.~H., {Weinmann}, S.~M., \& {Kang}, X. 2008, \mnras, 387, 79

\bibitem[{{Verdugo} {et~al.}(2008){Verdugo}, {Ziegler}, \&
  {Gerken}}]{2008A&A...486....9V}
{Verdugo}, M., {Ziegler}, B.~L., \& {Gerken}, B. 2008, \aap, 486, 9

\bibitem[{{Wilman} {et~al.}(2005){Wilman}, {Balogh}, {Bower}, {Mulchaey},
  {Oemler}, {Carlberg}, {Morris}, \& {Whitaker}}]{2005MNRAS.358...71W}
{Wilman}, D.~J., {Balogh}, M.~L., {Bower}, R.~G., {Mulchaey}, J.~S., {Oemler},
  A., {Carlberg}, R.~G., {Morris}, S.~L., \& {Whitaker}, R.~J. 2005, \mnras,
  358, 71

\bibitem[{{Wilman} {et~al.}(2008)}]{2008ApJ...680.1009W}
{Wilman}, D.~J., et al. 2008, \apj, 680, 1009

\bibitem[{{Worthey}(1994)}]{1994ApJS...95..107W}
{Worthey}, G. 1994, \apjs, 95, 107

\bibitem[{{Yan} {et~al.}(2008)}]{2008arXiv0805.0004Y}
{Yan}, R., et al. 2008, ArXiv e-prints

\bibitem[{{Yee}(1991)}]{1991PASP..103..396Y}
{Yee}, H.~K.~C. 1991, \pasp, 103, 396

\bibitem[{{Yee} {et~al.}(1996){Yee}, {Ellingson}, \&
  {Carlberg}}]{1996ApJS..102..269Y}
{Yee}, H.~K.~C., {Ellingson}, E., \& {Carlberg}, R.~G. 1996, \apjs, 102, 269

\bibitem[{{Yee} {et~al.}(2007){Yee}, {Gladders}, {Gilbank}, {Majumdar},
  {Hoekstra}, \& {Ellingson}}]{2007ASPC..379..103Y}
{Yee}, H.~K.~C., {Gladders}, M.~D., {Gilbank}, D.~G., {Majumdar}, S.,
  {Hoekstra}, H., \& {Ellingson}, E. 2007, in Astronomical Society of the
  Pacific Conference Series, Vol. 379, Cosmic Frontiers, ed. N.~{Metcalfe} \&
  T.~{Shanks}, 103--+

\bibitem[{{Yee} {et~al.}(2000)}]{2000ApJS..129..475Y}
{Yee}, H.~K.~C., et al. 2000, \apjs, 129, 475

\end{thebibliography}
        \begin{table}
        \caption{Properties of CNOC1 Follow-up Data Set \label{tb1}}
        \begin{tabular}{lllllll}
        \tableline\tableline \\
        Cluster & R.A.(J2000)& Dec.(J2000)& $z_{spec}$     & $m_{R_c,lim}$ & seeing$^a$ & note\\
        \tableline\\  
        MS0302+16   & 03:05:18.1 & 17:28:24.9 & 0.425 & 24.10 & 0.70 & classical \\
                      &            &            &       &       &     & not a CNOC1 cluster \\
        CL0303+17   & 03:06:18.7 & 17:18:03.0 & 0.418 &       &     & S-E field \\
                      &            &            &       &       &     & not a CNOC1 cluster \\
        MS0440+02     & 04:43:09.9 & 02:10:21.4 & 0.197 & 23.71 & 0.58 & classical \\
        MS0451+02     & 04:54:14.1 & 02:57:10.5 & 0.201 & 23.72 & 0.68 & classical \\
        MS0451-03     & 04:54:10.9 &-03:00:57.1 & 0.539 & 22.80 & 1.11 & classical \\
        MS0839+29     & 08:42:55.9 & 29:27:27.3 & 0.193 & 23.05 & 0.86 & classical \\
        MS0906+11     & 09:09:12.7 & 10:58:29.2 & 0.171 & 23.03 & 0.82 & classical \\
        MS1008-12     & 10:10:32.3 &-12:39:52.5 & 0.306 & 23.35 & 1.17 & classical \\
        MS1224+20     & 12:27:13.3 & 19:50:56.5 & 0.330 & 21.70 &      & queue, not use \\
        MS1231+15     & 12:33:55.4 & 15:25:59.1 & 0.235 & 23.80 & 0.82 & queue  \\
        Abell 1560    & 12:34:07.1 & 15:10:28.2 & 0.244 &       &      & S field \\
                      &            &            &       &       &     & not a CNOC1 cluster \\
        MS1358+62     & 13:59:50.5 & 62:31:04.5 & 0.329 & 22.65 & 1.24 & classical \\
        MS1358+62S    & 13:59:38.0 & 62:18:47.0 & 0.329 &       &      & S field \\
                      &            &            &       &       &     & not a CNOC1 cluster \\
        MS1455+22     & 14:57:15.1 & 22:20:33.9 & 0.257 & 23.20 & 0.93 & queue \\
        MS1512+36     & 15:14:22.5 & 36:36:20.9 & 0.373 & 23.85 & 0.87 & queue  \\ 
        MS1621+26     & 16:23:35.5 & 26:34:15.9 & 0.428 & 23.35 & 0.93 & classical \\
        Abell 2390    & 21:53:36.8 & 17:41:43.7 & 0.228 & 24.05 & 0.72 & classical \\
        \tableline\tableline\\
        \tablenotetext{a}{in $R_c$-band frame in arcsecond}
        \end{tabular}
        \end{table}

        \begin{table}
	\caption{Exposure Times$^b$ and Filters for Each Cluster \label{tb2}}
        \begin{tabular}{lccccclcccc}
        \tableline\tableline \\
        Cluster & $B$   & $V$   & $R_c$ & $I$ & & Cluster & $B$ & $V$ & $R_c$ & $I$ \\
        \tableline
        MS0302+16 & 92  & 20    & 16    & 16 &  & MS1224+20 & 69        & 12    & 12    & 12 \\
        MS0440+02 & 20  & 7     & 7     & 7  &  & MS1231+15 & 20        & 7     & 7     & 7 \\
        MS0451+02 & 20  & 7     & 7     & 7 &  & MS1358+62 & 69         & 15    & 10.5  & 15 \\
        MS0451-03 & 60  & 69    & 24    & 24 &  & MS1455+22 & 20        & 7     & 7     & 7 \\
        MS0839+29 & 20  & 7     & 7     & 7 &  & MS1512+36      & 23    & 15    & 14    & 14 \\
        MS0906+11 & 40  & 7     & 7     & 14 &  & MS1621+26 & 20        & 12    & 16    & 10 \\
        MS1008-12 & 44  & 7     & 14    & 7  &  & Abell 2390   &  15 & 10 & 10 & 12 \\
        \tableline\tableline \\
        \tablenotetext{b}{in minutes}
        \end{tabular}
        \end{table}

	\begin{table}
        \caption{Properties of pFoF Determined CNOC1 Clusters ($D0_{xy}=0.25 Mpc$) \label{Tmainbody}}
        \begin{tabular}{lcccccc}
        \tableline\tableline\\
Cluster & R.A.(J2000) & Dec.(J2000) & $z_{pFoF,cl}$ & $N_{gal}$ & $R_{200}^c$ & $f_{red}$\\
\tableline\\
MS0302+16 & 03:05:18.25 & 17:28:29.5 & $0.4488\pm 0.0095$ & $27.4\pm 2.4 $ & 1.20 & $0.698\pm 0.115$ \\
CL0303+17 & 03:06:19.69 & 17:18:29.9 & $0.4778\pm 0.0045$ & $135.0\pm 5.0 $ & 2.05 & $0.624\pm 0.053$ \\
MS0440+02 & 04:43:09.95 & 02:10:21.6 & $0.1989\pm 0.0032$ & $29.5\pm 2.3 $ & 1.24 & $0.922\pm 0.057$ \\
MS0451+02 & 04:54:13.85 & 02:57:07.4 & $0.1998\pm 0.0005$ & $208.5\pm 12.6 $ & 1.97 & $0.947\pm 0.080$ \\
MS0451-03 & 04:54:12.60 & -00:00:48.4 & $0.5260\pm 0.0041$ & $222.1\pm 12.9 $ & 2.07 & $0.357\pm 0.051$ \\
MS0839+29 & 08:42:56.06 & 29:27:40.8 & $0.1996\pm 0.0018$ & $70.4\pm 2.7 $ & 1.60 & $0.831\pm 0.056$ \\
MS0906+11 & 09:09:07.57 & 10:57:55.1 & $0.2060\pm 0.0042$ & $36.5\pm 2.8 $ & 1.79$^d$ & $0.939\pm 0.045$ \\
MS1008-12 & 10:10:32.45 & -00:39:52.2 & $0.3300\pm 0.0051$ & $57.4\pm 3.6 $ & 1.94 & $0.933\pm 0.043$ \\
MS1231+15 & 12:33:54.51 & 15:26:59.1 & $0.2348\pm 0.0034$ & $39.0\pm 2.5 $ & 1.30 & $0.907\pm 0.055$ \\
Abell 1560 & 12:34:14.46 & 15:13:40.1 & $0.2486\pm 0.0029$ & $60.8\pm 3.4 $ & 1.57 & $0.942\pm 0.041$ \\
MS1358+62 & 13:59:50.33 & 62:31:00.7 & $0.3390\pm 0.0037$ & $65.3\pm 4.2 $ & 1.64 & $0.959\pm 0.029$ \\
MS1358+62S & 13:59:34.04 & 62:19:05.2 & $0.3347\pm 0.0040$ & $53.5\pm 3.7 $ & 1.50 & $0.932\pm 0.043$ \\
MS1455+22 & 14:57:11.66 & 22:19:54.2 & $0.2386\pm 0.0051$ & $33.1\pm 2.4 $ & 2.24 & $0.903\pm 0.067$ \\
MS1512+36 & 15:14:22.43 & 36:36:23.1 & $0.3771\pm 0.0093$ & $15.6\pm 2.0 $ & 1.21 & $0.915\pm 0.067$ \\
MS1621+26 & 16:23:35.48 & 26:34:52.1 & $0.4315\pm 0.0033$ & $42.5\pm 3.0 $ & 1.39 & $0.884\pm 0.061$ \\
Abell 2390 & 21:53:37.06 & 17:41:17.1 & $0.2202\pm 0.0012$ & $126.7\pm 4.3 $ & 2.16 & $0.927\pm 0.036$ \\
        \tableline\\
        \tablenotetext{c}{in \it Mpc\rm, $h$=0.7}
        \tablenotetext{d}{using the single peak in the velocity dispersion in \cite{1999ApJ...527..561B}}
        \end{tabular}
\end{table}


\clearpage

\clearpage


\end{document}